\documentclass[aps,prb,times,twocolumn,superscriptaddress,showpacs,a4paper]{revtex4-1}

\usepackage{colortbl}
\usepackage{amsfonts,amsmath,amssymb}
\usepackage[dvipdfmx]{graphicx}
\usepackage{hyperref,tikz}
\usepackage{tabularx,hhline}
\usepackage{float} 
\usepackage{dcolumn,stmaryrd}
\usepackage{ulem}

\begin{document}

\title{
When opposites repel: from metastability to extended chiral spin textures in \\ spin ice with short-range topological-defect interactions
}

\author{M. Udagawa}
\affiliation{Department of Physics, Gakushuin University, Mejiro, Toshima-ku, Tokyo 171-8588, Japan}
\email{masafumi.udagawa@gakushuin.ac.jp}

\author{L.D.C. Jaubert}
\affiliation{Okinawa Institute of Science and Technology Graduate University,
Onna-son, Okinawa 904-0495, Japan}

\author{C. Castelnovo}
\affiliation{TCM group, Cavendish Laboratory, University of Cambridge, Cambridge, CB3 0HE, United Kingdom}

\author{R. Moessner}
\affiliation{Max Planck Institute for the Physics of Complex Systems, N\"{o}thnitzer Str.~38, D-01187 Dresden, Germany}

\date{\today}

\begin{abstract}
We study the interplay of topological bottlenecks and energetic barriers to equilibration in a Coulomb spin liquid where a short-range energetic coupling between defects charged under an emergent gauge field supplements their entropic long-range Coulomb interaction. This work is motivated by the prevalence of memory effects observed across a wide range of geometrically frustrated magnetic materials, possibly including the spontaneous Hall effect observed in Pr$_{2}$Ir$_{2}$O$_{7}$. Our model is canonical spin-ice model on the pyrochlore lattice, where farther-neighbour spin couplings  give rise to a nearest-neighbor interaction between  topological defects which can easily be chosen to be {\it ``unnatural"} or not, i.e.\ attractive or repulsive between defects of equal gauge charge. Among the novel features of this model are the following. 
After applying a  field quench, a rich dynamical approach to equilibrium emerges, dominated by multi-scale energy barriers responsible for long-lived magnetization plateaux. These even allow for the metastability of a ``fragmented'' spin liquid, an elusive regime where partial order co-exists with a spin liquid. Perhaps most strikingly, the attraction produces clusters of defects whose stability is due to a combination of energetic barriers for their break-up and proximity of opposite charges along with an entropic barrier generated by the topological requirement of annihilating a defect only together with an oppositely charged counterpart. These clusters may take the form of a ``jellyfish'' spin texture, comprising an arrangement of same-sign gauge-charges, centered on a hexagonal ring with branches of arbitrary length. The  ring carries a clockwise or counterclockwise circular flow of magnetisation. This emergent toroidal degrees of freedom provides a possibility for time reversal symmetry breaking with possible relevance to the spontaneous Hall effect observed in Pr$_2$Ir$_2$O$_7$.
\end{abstract}
\pacs{75.10.Hk, 75.40.Gb, 75.60.Nt}
\maketitle

\section{Introduction}
In spin glasses, frustration is an essential ingredient of the glassiness~\cite{EdwardsAnderson1975}, responsible for nonequilibrium phenomena such as memory effects. But even in disorder-free systems, frustration, now of geometrical origin, imposes considerable constraints on the kinetics, responsible for a diverse range of unconventional dynamical behavior. The interplay between frustration and spin dynamics has been a long-standing source of questions in condensed matter and statistical mechanics~\cite{Bouchaud97a}. For example, the peculiar dynamical response observed in triangular-based organic systems~\cite{Kagawa2013} has been intensively discussed recently in terms of geometrical frustration.

Indeed, while geometrically frustrated magnets lack disorder to generate a rugged energy landscape, they have topology as a new ingredient for the generation of slow dynamics: in the simplest terms, gauge-charged topological defects cannot spontaneously disappear but rather only pair-annihilate with an oppositely charged partner. This provides connections to the physics of reaction-diffusion systems, as well as to the coarsening literature, and more generally to the study of kinetically constrained models~\cite{Toussaint83,Stinchcombe01,Ritort03}. 

While such topological constraints provide hard conditions on the kinds of allowed dynamical processes, they can be supplemented by soft yet important, non-topological energetic considerations. Adding short-range interactions between topological defects can give rise to a rich phenomenology which has not yet been thoroughly or systematically explored--for a case study of quantum Hall physics and the links to different types of superconductivity, see Ref.~[\onlinecite{Parameswaran12}]. Our work aims to add considerations of non-equilibrium dynamics to this intriguing set of phenomena.

A well-studied and experimentally relevant case in point is provided by spin ice, a canonical model and family of materials with strong geometrical frustration\cite{Harris97,Ramirez99}, whose equilibrium properties are well explored. The ground state of the spin ice model has macroscopic degeneracy, and its spatial structure can be described by a free emergent gauge-field arising from a divergence-free condition on the spin density imposed energetically. Excitations out of this ground-state ensemble are analogues of magnetic monopoles~\cite{Castelnovo2008}, interacting together via an effective magnetic Coulomb potential.

From a dynamical point of view, the divergence-free condition and interaction between monopoles imposes strong microscopic kinematical constraints on the motion of spins. Among other consequences, the magnetic relaxation time diverges at low temperature in Dy$_{2}$Ti$_{2}$O$_{7}$ and Ho$_{2}$Ti$_{2}$O$_{7}$ spin-ice materials, as measured by a variety of experimental probes, such as AC-magnetic susceptibility~\cite{Matsuhira2001,Snyder2001,Quilliam2011}, thermal transport~\cite{Klemke2011}, neutron spin echo~\cite{Ehlers2003}, neutron scattering~\cite{Clancy2009} and muon spectroscopy~\cite{Lago2007}. This spin freezing is due to the rarefaction of defects~\cite{Jaubert2009,Jaubert2011}, mediated by impurities and surface effects~\cite{Revell2013}. Indeed, defects play a fundamental role in facilitating spin dynamics~\cite{Ryzhkin05}. In fact, spin ice has been shown to be a fantastic framework for unconventional nonequilibrium physics~\cite{Mostame2014}, where monopoles can form non-contractible pairs~\cite{Castelnovo2010} (see below).

In this context of anomalous spin-ice dynamics, the metallic spin-ice material, Pr$_2$Ir$_2$O$_7$, has recently attracted considerable attention. In this compound, the Ir $5d$ conduction electrons interact with a magnetic `spin-ice' texture formed by the localized Pr$^{3+}$ moments. The enhanced spin-ice correlations induce anomalous scattering of conduction electrons\cite{Nakatsuji06,Machida07,Machida10,Balicas11}, and the resultant unusual transport properties have generated considerable theoretical interest\cite{Udagawa10,Udagawa13,Moon2013,Flint2013,Chern2013,Lee2013}. For instance, the Hall conductivity shows non-monotonic magnetic field dependence, implying that the Hall response is dominated by the topological Hall effect due to the scattering of itinerant electrons from spin triplets with finite spin scalar chirality\cite{Machida07,Udagawa13}.

Even more striking is the so-called spontaneous Hall effect observed in this system, where a finite Hall response is obtained with neither magnetic field nor spontaneous magnetization. This implies the formation of exotic states with broken time-reversal symmetry, but without ferromagnetic order. To describe the experimental setting more precisely, the system is initially placed in a magnetic field of 7 Tesla in the $[111]$ direction. After the magnetic field is removed, magnetization relaxes to zero, while the finite Hall signal remains~\cite{Machida07}.
This phenomenon has invited several interpretations, such as chiral spin liquid formation~\cite{Machida07}. 

Possibly, this spontaneous Hall effect may alternatively be attributed to the nature of low-energy excitations. The initial state obviously breaks time-reversal symmetry under a magnetic field, and naturally shows a finite Hall response. Accordingly, if there exist non-magnetic excitations which somehow encode the information of time-reversal symmetry breaking, then the population of these excitations reflect the broken time-reversal symmetry of the initial state, and the Hall response might persist for a period after the field quench.
Moreover, if the excitation has a long lifetime, and do not relax within observable time scales, even the ``steady" state may retain a finite Hall response.

In fact, the existence of long-living composite excitations is known in dipolar spin ice. Non-contractible pairs of monopoles are formed through the attractive interaction between monopoles, and they exhibit long lifetime\cite{Castelnovo2010}, as their pair annihilation can only proceed across an energy barrier.
The formation of composite excitations may also occur in metallic spin ice, where the magnetic moments interact through an RKKY-type interaction, where the farther-neighbor interaction brings about effective interactions between topological defects. It is interesting to examine how the interaction between topological defects affects the macroscopic dynamics of the system and the emergence of novel composite excitations. 

\begin{figure}[t]
\begin{center}
\includegraphics[width=0.4\textwidth]{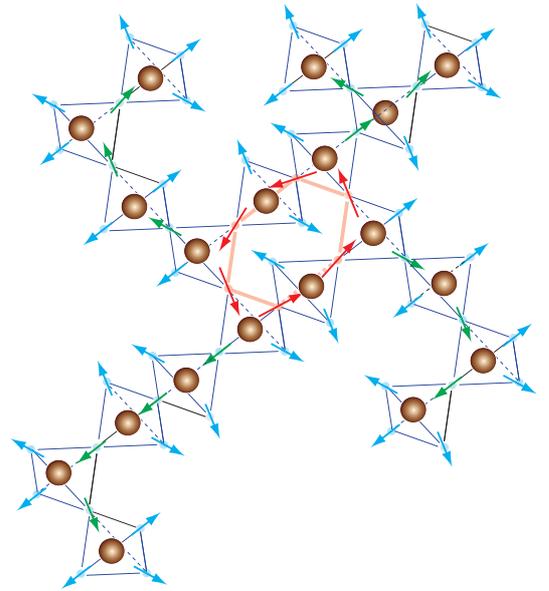}
\end{center}
\caption{\label{schematic_jellyfish} 
(color online). A spin configuration of the jellyfish structure composed of positive gauge-charges. The jellyfish structure has a central ring composed of minimally 6 tetrahedra from which branches of same-sign charges emerge. The loop of spins running along the central ring possesses a chirality whose value (clockwise or counterclockwise) is independent of the sign of the charges. In the branches however, there is an unidirectional flux of magnetization which is imposed by the sign of the charges. In this example, the magnetization flux goes away from the central ring, as can be seen from following the minority spins of each tetrahedron (shown in green).
}
\end{figure}

\section{Model and summary of main results}

We study the $J_1-J_2-J_3$ spin-ice model, which extends the nearest-neighbor spin-ice model, by adding second-neighbor ($J_2$) and third-neighbor ($J_3$) couplings. In this work, we focus on the case $J_2=J_3(\equiv J)$, setting $J_1=1$. This model can elegantly be rewritten in terms of gauge-charged degrees of freedom~\cite{Ishizuka2013}. The spin-ice degeneracy is preserved and remains in the ground state of the model for an extended region of couplings $J\in[-1/2:1/4]$. While the ground state belongs to the vacuum sector of a Coulomb phase, its excitations are described as positive and negative charges, which are called monopoles in the context of dipolar spin ice. Interestingly, 
an emergent part of the interaction between charges for $J \neq 0$ is short-ranged, and its sign is tunable by changing that of $J$. Quasi-particles with the same topological charge attract (repel) for $J>0 (<0)$.

Given these basic properties of the model, we first summarize the main results of our analysis. Firstly, let us focus on the region of $J>0$. This region is unusual in the sense that ``like" charges, \textit{i.e.} quasi-particles having the same gauge charge, attract, in contrast to, e.g., monopoles in dipolar spin ice. This property immediately leads to the existence of collective excitations composed of like charges, which we pictorially name ``jellyfish". A jellyfish excitation is an extended structure consisting of a central ring with an arbitrary numbers of branches attached. Its excitation energy decreases with  increasing interaction $J$, vanishing at the critical point, $J=1/4$. In addition to the energetic stability mentioned above, the jellyfish excitation enjoys ``kinematic stability", i.e. the clustering of same charges decreases the opportunities of pair-annihilation, which occurs only between opposite charges. The behavior of these charges is the subject of our analysis of their stochastic dynamics.

The dynamical bottleneck caused by the jellyfish excitations leads to the possibility of interesting memory effects, including an emergent chiral degree of freedom which carries zero magnetization. Indeed, the ring part of jellyfish carries a clockwise or counterclockwise circular flow of magnetization, which gives it a well-defined toroidal moment. The stability of jellyfish excitations implies a slow relaxation of these toroidal moments. Accordingly, once the system is subject to a perturbation which breaks time-reversal symmetry, e.g. a magnetic field, this symmetry breaking can in principle persist for a long time, even after relaxation of the magnetization. In the context of a magnetic field quench, this suggests that a signal of broken time-reversal symmetry can be detected a long time after the removal of the magnetic field, e.g. through the Hall response. This provides a possible scenario for the mysterious spontaneous Hall effect observed in Pr$_2$Ir$_2$O$_7$.

As well as in the dynamical properties, the jellyfish excitations leave their fingerprints in thermodynamic quantities, which we reveal by combining Monte Carlo simulations and the analytical Bethe approximation. Approaching the critical point, $J=1/4$, the charge density exhibits a non-monotonic temperature dependence, reflecting the softening of jellyfish excitations. An immediate consequence of the softening is found in the entropy. At $J=1/4$, the zero-energy jellyfish add contribution to the ground state degeneracy of the spin ice manifold, enhancing the residual value of the entropy. The crossover behavior at the energy scale of jellyfish excitation gives a clear signature in the magnetic susceptibility, $\chi_{z}$. Near $J=1/4$, $\chi_{z}T$ approaches its high temperature value as the population of jellyfish excitation grows, instead of showing a monotonic Curie-law crossover as in ideal spin ice\cite{Jaubert2013}.
Moreover, in addition to these thermodynamic quantities, the jellyfish affects the magnetic structure factor considerably. Instead of pinch point singularities, which serve as an icon of the vacuum of the Coulomb phase, a ``half-moon" pattern appears in the magnetic structure factor. The detection of this pattern through quasi-elastic neutron scattering can serve as a signature of jellyfish structures.

While the region of $J>0$ exhibits rich behavior both dynamically and thermodynamically, fertile non-equilibrium behavior is also observed for $J<0$, where opposite charges attract as is ``normal". For small negative $J$, the magnetization shows markedly slow relaxation, compared to the  charge density. The magnetization takes a constant value over a wide time range. The formation of this magnetization plateau is associated with the exhaustion of charges, and the time scale of its termination can be understood from the ``vacuum creation" of pairs of gauge charges.

Approaching $J=-1/2$, the dynamics changes drastically and becomes dominated by the influence of double charges. As a result, charge relaxation becomes extremely slow, while the magnetization decays rather quickly. The system finally forms the co-called fragmented Coulomb spin liquid (FCSL)~\cite{Borzi2013,Brooks2014,Jaubert2015}. In the FCSL charges are long-range ordered but the spin texture remains disordered, extensively degenerate and described by a Coulomb gauge theory. We shall here make the following distinction. A configuration in the Coulomb phase of spin ice is entirely covered with two-in two-out tetrahedra, while a FCSL configuration is alternatively covered by 3-in 1-out and 3-out 1-in tetrahedra: the four nearest-neighbours of a 3-in 1-out tetrahedron are 3-out 1-in tetrahedra, and vice-versa. The FCSL has been predicted theoretically~\cite{MoellerMoessner2009,Chern2011} and observed experimentally~\cite{Arnalds12,Zhang2013,Anghinolfi15} in nano-lithographic artificial kagome ice whose geometry prevents the existence of a charge-free Coulomb phase~\cite{NisoliMoessnerSchiffer2013}. But in three dimensions, it has so far only been partially stabilized at equilibrium in the spin-ice model with dipolar interactions~\cite{Guruciaga2014}, or requires four-body interactions~\cite{Jaubert2015} or the suppression of double charges~\cite{Borzi2013,Brooks2014}. The nonequilibrium magnetic-field quench proposed here provides a promising tool to realize this state and demonstrates the possibility of engineering a macroscopic state via nonequilibrium techniques~\cite{Paulsen2014}.\\

The organization of this paper is as follows. In section~\ref{sec:model}, we present the general form of the spin model with first, second and third nearest-neighbor interactions. We rewrite this model with gauge charge degrees of freedom, which we name nearest-neighbor dumbbell model, for a specific line of parameters. The corresponding phase diagram is discussed in section~\ref{section_phase_diagram}. The nonequilibrium properties of the model for a field quench are extensively investigated in section~\ref{noneq}. These results identify a promising point, $J = 1/4$, on the phase diagram with an enhanced emergent degeneracy. We present a thorough analysis of the equilibrium properties around this point in section~\ref{thermodynamics}. Consequences and future directions of research are  discussed in the conclusions.

\begin{figure}[t]
\begin{center}
\includegraphics[width=0.45\textwidth]{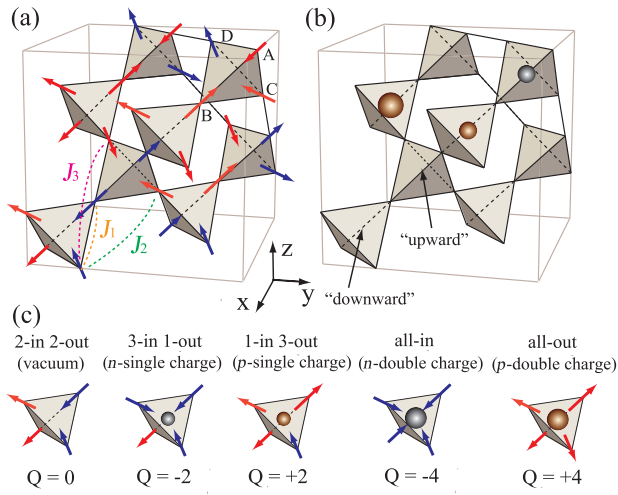}
\end{center}
\caption{\label{model_convention} 
(color online). (a) An example of spin configuration. The spin pairs connected by interactions, $J_1$, $J_2$ and $J_3$ are shown with dashed lines. $J_{3}$ does not include interactions across hexagons.
(b) The charge distribution corresponding to the spin configuration shown in (a). We also show here the two classes of tetrahedra, which we call ``upward" and ``downward", according to their orientations. (c) Examples of spin configurations are shown, corresponding to each value of tetrahedral charge. For simplicity, we refer to the tetrahedra with $Q=0, \pm2$ and $\pm4$ as
vacuum, positive(negative) single charge, and positive(negative) double charge, respectively.
}
\end{figure}

\section{The model}
\label{sec:model}

\subsection{$J_1-J_2-J_3$ spin ice model}
\label{J1J2J3spinicemodel}
We shall consider a set of Ising moments at each site $i$ of a pyrochlore lattice, ${\mathbf S}_i=\eta_i{\mathbf d}_i$ where ${\mathbf d}_i$ defines the local easy axis, and $\eta_i=\pm1$ is the Ising variable. The four sublattices of the pyrochlore lattice are defined in Fig.~\ref{model_convention}, with the following easy-axes:
\begin{eqnarray}
\mathbf{d}_{\{A,B,C,D\}}=\frac{1}{\sqrt{3}}\left\{
\begin{pmatrix}1\\1\\1\end{pmatrix},
\begin{pmatrix}1\\-1\\-1\end{pmatrix},
\begin{pmatrix}-1\\1\\-1\end{pmatrix},
\begin{pmatrix}-1\\-1\\1\end{pmatrix}
\right\}
\label{eq:di}
\end{eqnarray}
By definition, the vectors ${\mathbf d}_{i}$ point outward (resp. inward) the so-called ``upward" (resp. ``downward") tetrahedra [see Fig.~\ref{model_convention}.(b)]. We define the Hamiltonian of the $J_1-J_2-J_3$ spin ice model as
\begin{align}
\mathcal{H} &= \tilde{J}_1\sum_{n.n.}{\mathbf S}_i\cdot{\mathbf S}_j + \tilde{J}_2\sum_{2nd.}{\mathbf S}_i\cdot{\mathbf S}_j + \tilde{J}_3\sum_{3rd.}{\mathbf S}_i\cdot{\mathbf S}_j,\nonumber\\
&= J_1\sum_{n.n.}\eta_i\eta_j + J_2\sum_{2nd.}\eta_i\eta_j + J_3\sum_{3rd.}\eta_i\eta_j,
\label{spinHamiltonian}
\end{align}
including successively the first, second and third nearest-neighbor coupling [see Fig.~\ref{model_convention}]. The nearest-neighbor interaction leads to the spin-ice model with the 2-in 2-out Coulomb phase dominating the physics below a temperature of order $J_1$. Hereafter, we set $J_1=1$, as a unit of energy and temperature. Meanwhile the second and third terms are expected to considerably alter the qualitative nature of the ground state. However, as we shall discuss in this paper, if $J_2=J_3\equiv J$, the spin-ice ground state is preserved for a broad range of values of $J$.

One of our motivations to consider the $J_1-J_2-J_3$ spin-ice model comes from its application as an effective model for spin degrees of freedom in pyrochlore conductors. In a group of pyrochlore oxides, such as Pr$_2$Ir$_2$O$_7$, the rare-earth moments behave as Ising moments and interact with each other through the RKKY interaction mediated by itinerant electrons originated from, e.g., $d$ orbitals of transition metal ions. Usually, the RKKY interaction is long-ranged, however, its fast-decaying and oscillating nature makes it possible to approximate it with a short-range model. Indeed, the phase diagram of the Ising Kondo lattice model, where the effect of itinerant electrons is fully taken into account, is quite well reproduced by the $J_1-J_2-J_3$ spin-ice model\cite{Ishizuka2014}.

\subsection{Gauge-charge representation: $J_2=J_3=J$}
As mentioned in the previous section, the $J_1-J_2-J_3$ spin-ice model completely preserves the two-in two-out degeneracy, as long as $J_2=J_3$ is satisfied. To see this, it is convenient to rewrite the Hamiltonian (\ref{spinHamiltonian}) in terms of charge degrees of freedom, $Q_p$, as was first done by Ishizuka et al. \cite{Ishizuka2013}. For a tetrahedron $p$ composed of the sites, $p1, p2, p3$ and $p4$, the charge is defined as
\begin{eqnarray}
Q_p = \zeta_p(\eta_{p1} + \eta_{p2} + \eta_{p3} + \eta_{p4}),
\end{eqnarray}
where $\zeta_p=+1 (-1)$, if $p$ corresponds to an upward (downward) tetrahedron. Please note that the sign convention used here is opposite to the one of magnetic monopoles in the dumbbell model~\cite{Castelnovo2008}. This is an arbitrary choice, possible because while the sign of magnetic monopoles is a physical quantity imposed by the inherent nature of magnetic dipolar interactions, the objects considered here are topological charges whose convention is for us to select. If the configuration of tetrahedron $p$ is two-in two-out, then $Q_p=0$, while $Q_p=\pm2$ or $\pm4$ for single and double charges respectively [see Fig.~\ref{model_convention}]. Then, if we rewrite Hamiltonian (\ref{spinHamiltonian}) as a function of charge degrees of freedom for $J_2=J_3$, we obtain the nearest-neighbour dumbbell model:
\begin{eqnarray}
\mathcal{H} = \Bigl(\frac{1}{2} - J\Bigr)\sum_{p}Q_p^2 - J\sum_{\langle p,q\rangle}Q_p\,Q_q.
\label{chargeHamiltonian}
\end{eqnarray}
The details of the derivation are given in appendix \ref{derive_charge}. The two terms in the Hamiltonian (\ref{chargeHamiltonian}) allow for simple interpretations: the creation cost and the nearest-neighbor interaction between charges. The two-in two-out degeneracy is trivially preserved ($Q_{p}=0$ everywhere), and, as we shall see, even remains the ground state of Hamiltonian (\ref{chargeHamiltonian}) for small values of $|J|$. In other words, in order to lift the degeneracy of spin ice with a short-ranged perturbation, the relevant energy scale is $|J_2-J_3|$, rather than $|J_2|$ or $|J_3|$.

The main difference with the standard dumbbell model of magnetic monopoles in spin ice~\cite{Castelnovo2008} is that the interaction between charges is now nearest-neighbour and can be either attractive or repulsive between same-sign charges. Also, our model is a one-parameter problem, $J$, which means that the creation cost of charges is directly linked to the strength of the interaction.


\section{Phase diagram at equilibrium}
\label{section_phase_diagram}

\begin{figure}[h]
\begin{center}
\includegraphics[width=0.5\textwidth]{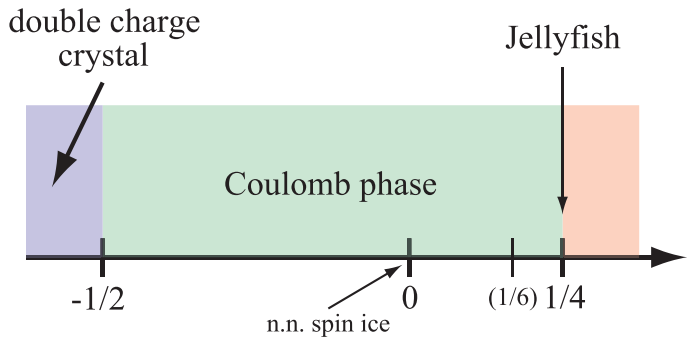}
\end{center}
\caption{\label{Basic} 
(color online). Ground-state phase diagram of Hamiltonian (\ref{chargeHamiltonian}) as a function of $J$. This corresponds to the $J_1-J_2-J_3$ spin-ice model for $J_{2}=J_{3}=J$, and $J_1=1$. The Coulomb phase is stabilized over a broad region ($-1/2<J<1/4$). For $J<-1/2$, it gives way to the antiferromagnetic phase made of alternating all-in and all-out tetrahedra (double charges). At $J=1/4$, the jellyfish point discussed in section~\ref{thermodynamics} appears.
}
\end{figure}

Before going into the analysis of the dynamics, we shall discuss the equilibrium phase diagram of this model. For $J_2=J_3=J$, the phase diagram is divided into three regions, as shown in Fig.~\ref{Basic}. At $J=0$, the model is precisely the nearest-neighbor spin ice model, and its ground state is the well-known Coulomb phase where all tetrahedra are in the two-in two-out configurations [see \textit{e.g. }the pedagogical review by Chris Henley~[\onlinecite{Henley2010}]]. Remarkably, the nearest-neighbor and the present spin ice models are projectively equivalent: the extensive degeneracy of the Coulomb phase is preserved for finite $J$, and it remains ground state of the model for a wide region of parameters, $-1/2<J<1/4$.\\

To understand the robustness of the Coulomb phase, we rewrite the charge Hamiltonian given in equation (\ref{chargeHamiltonian}) as
\begin{eqnarray}
\mathcal{H} = \Bigl(\frac{1}{2}+J\Bigr)\sum_p Q_p^2 - \frac{J}{2}\sum_{\langle p,q\rangle}(Q_p + Q_q)^2.
\label{charge1}
\end{eqnarray}
Since both sums are non-negative, one can immediately conclude that for $-1/2<J\leq0$ the ground-state configurations satisfy $Q_p=0$ for all tetrahedra $p$, those of the Coulomb phase.

Meanwhile, for $J<-1/2$, the first term of Eq.~(\ref{charge1}) becomes negative and is now minimized if $|Q_p|$ is maximized, i.e., $Q_p=\pm4$ for all $p$. At the same time, the second term is minimized if $Q_p + Q_q=0$. These two conditions are simultaneously achieved, if and only if $Q_p=+4 (-4)$ for all upward (resp. downward) tetrahedra. In terms of spins, this corresponds to all-in / all-out configurations, which can be seen as a double charge crystal (zinc-blende structure).

Right at $J=-1/2$, the creation cost of charges in Hamiltonian~(\ref{charge1}) vanishes, making the double charge crystal degenerate with the Coulomb phase. Moreover, charge crystals made of alternating charges with $Q_p=+2$ and $-2$ on upward and downward tetrahedra respectively also minimize the energy [see an example of spin configuration in Fig.~\ref{monopolecrystal}]. This corresponds to the fragmented Coulomb spin liquid~\cite{Borzi2013,Brooks2014} where long-range charge order co-exists with extensive spin degeneracy.

As for the other side of the phase diagram, $J>0$, the limit of stability of the Coulomb phase is more subtle. If one rewrites the Hamiltonian(\ref{chargeHamiltonian}) as
\begin{eqnarray}
\mathcal{H} = \Bigl(\frac{1}{2}-3J\Bigr)\sum_p Q_p^2 + \frac{J}{2}\sum_{\langle p,q\rangle}(Q_p - Q_q)^2,
\label{charge2}
\end{eqnarray}
it guarantees that the ground state satisfies $Q_p=0$ for all tetrahedra $p$, at least up to $J=1/6$. The interaction term [see Eq.~(\ref{chargeHamiltonian})] clearly favors same-sign nearest neighbours, a condition that causes topological problems. First of all, charge neutrality must be preserved over the entire system: there must be as many positive as negative charges, which means that not all bonds can satisfy the condition $(Q_p - Q_q)=0$ for nearest neighbours $p$ and $q$. Furthermore, when one tries to cover the pyrochlore lattice with only, say, positive charges, one quickly encounters steric problems imposing the inclusion of vacuum or negative charges, even at a short length scale.

According to equation~(\ref{chargeHamiltonian}), it is relatively easy to see that there are only two kinds of spin clusters that minimize the interaction energy [Fig.~\ref{MinimalCluster}]:
\begin{itemize}
\item a double charge surrounded by four same-sign single charges,
\item a closed ring of same-sign single charges (this will give rise to the jellyfish structure discussed in detail in section~\ref{thermodynamics}).
\end{itemize}
In both cases, one gets locally for the tetrahedra which are part of these minimal clusters
\begin{eqnarray}
\sum_{p \in {\rm cluster}} Q_p^2 = \sum_{\langle p,q\rangle \in {\rm cluster}}Q_p Q_q.
\label{eq:eq}
\end{eqnarray}
Furthermore, it is possible to attach same-sign single charges to these minimal clusters, but no more double charges. Indeed if one takes the example of negative charges, then all spins on the outskirt of the cluster point outwards, which strictly forbids to attach any more double-negative charges. The addition of same-sign single charges to these minimal clusters also preserves the equality of equation~(\ref{eq:eq}). But as we have just discussed, not all $\langle p,q\rangle$ bonds can satisfy the condition $(Q_p - Q_q)=0$. It means that these clusters cannot cover the entire system. When summed over all tetrahedra, one gets
\begin{eqnarray}
\sum_p Q_p^2 \geq \sum_{\langle p,q\rangle}Q_p Q_q.
\label{eq:ineq}
\end{eqnarray}
According to Hamiltonian~(\ref{chargeHamiltonian}), it means that the total energy $E$ of any spin configuration for $J>0$ has a strict lower bound:
\begin{eqnarray}
E \geq \Bigl(\frac{1}{2} - 2J\Bigr)\sum_{p}Q_p^2.
\label{lowerbound}
\end{eqnarray}
For $0<J < 1/4$, the two-in two-out Coulomb phase thus remains the ground state of the system. Since we have shown the existence of configurations where the inequality of equation~(\ref{lowerbound}) is a strict equality, the Coulomb phase is not stable anymore for $J>1/4$. As for the physics at $J=1/4$, this deserves a dedicated discussion in section~\ref{thermodynamics}.


\section{Nonequilibrium dynamics via field quenches}
\label{noneq}
In this section, we investigate the stochastic dynamics of the model (\ref{spinHamiltonian}). To characterize the system, we focus on the single charge density $\rho_1$, and magnetization along $[111]$ axis. $\rho_1$ is defined as the sum of positive and negative single charges, divided by the number of tetrahedra, so that the maximal value of $\rho_1$ is normalized as 1. $M$ is also normalized to be $1$, if the system is in a saturation field in $[111]$ direction.

\subsection{Nearest-neighbor spin-ice model}
\label{nearest_neighbor_spin_ice}

\subsubsection{Temperature quench as a test}
\begin{figure}[h]
\begin{center}
\includegraphics[width=0.5\textwidth]{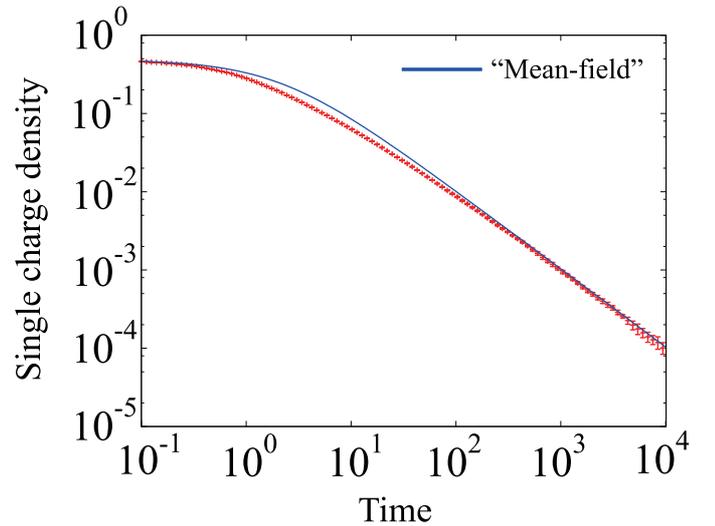}
\end{center}
\caption{\label{Tquench} 
(color online). Time evolution of the single charge density for the nearest-neighbor spin ice model ($J=0$) after a temperature quench from $T=10$ to $T=0$ starting at $t=0$. The solid curve is the mean-field solution of a creation-annihilation reaction equation: $\rho_1(t) = \dfrac{\rho_i}{1+ \mathcal{K}\rho_it}$, where $\mathcal{K}=\dfrac{9\sqrt{3}}{16}$ in agreement with Ref.~[\onlinecite{Castelnovo2010}].
}
\end{figure}

Firstly, we review the results for nearest-neighbor spin ice: $J_2=J_3=0$ in Fig.~\ref{Tquench}. Here, the temperature is initially set at $T=10$, then suddenly quenched to $T=0$ at time $t=0$. $J_{1}=1$ as reference energy. In Fig.~\ref{Tquench}, we plot the time dependence of the charge density. In this setting, a finite charge density of $\rho_i=0.487$ is thermally excited at $t=0$. After the temperature quench, charges start to annihilate in pairs, and their density decreases monotonically. Their density time dependence can be described by a simple reaction-type equation~\cite{Castelnovo2010}.

\subsubsection{Numerical setup for field quenches}
But temperature is not the only variable that can be used for a quench. In an anisotropic system such as spin ice, field quenches offer an alternative route for nonequilibrium phenomena, which are at the core of this paper.

\begin{figure}[h]
\begin{center}
\includegraphics[width=0.5\textwidth]{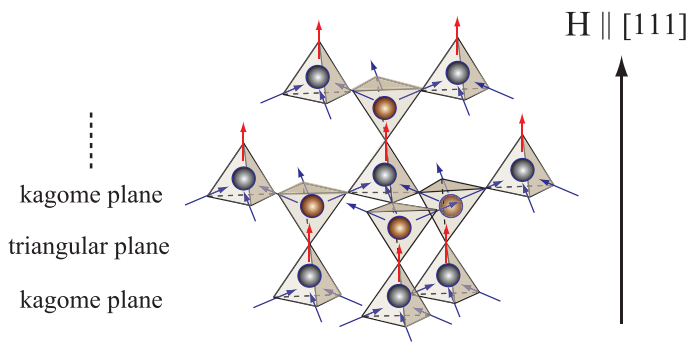}
\end{center}
\caption{\label{initialconfig111} 
(color online). The spin configuration under the $[111]$ saturation field. All the tetrahedra are occupied with 3-in/1-out or 1-in/3-out configurations. From the $[111]$ direction, the pyrochlore lattice can be regarded as an alternate stacking of kagome and triangular lattices. Please note that this perfectly aligned configuration belongs to the ensemble of fragmented Coulomb spin liquid states.
}
\end{figure}

{For this purpose, we supplement the Hamiltonian (\ref{spinHamiltonian}) with the Zeeman term,
\begin{eqnarray}
\mathcal{H}_{\rm Z} = -{\mathbf H}\cdot\sum_i{\mathbf S}_i.
\end{eqnarray}
}
In a field quench, the system is initially set under the magnetic field ${\mathbf H}\parallel[111]$, i.e. parallel to one of the easy axes, with $|{\mathbf H}|=100$, which can be practically regarded as a saturation field. Then, the field is removed suddenly at time $t=0$. At the initial stage, all the spins are aligned in the [111] direction, and all the tetrahedra are occupied with charges accordingly [see configuration in Fig.~\ref{initialconfig111}].

\begin{figure}[h]
\begin{center}
\includegraphics[width=0.5\textwidth]{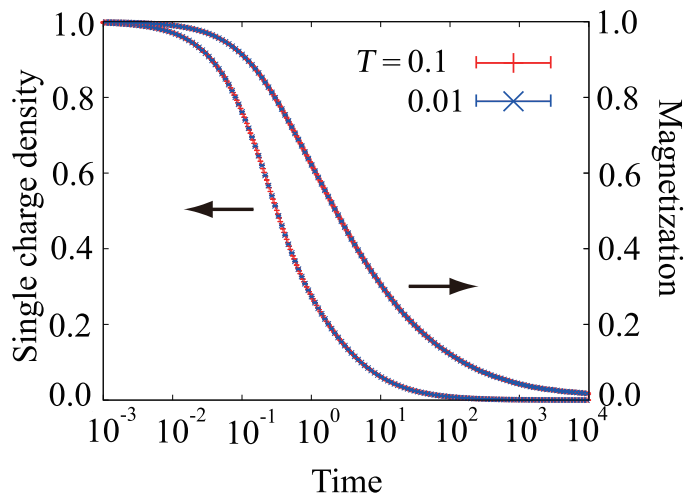}
\end{center}
\caption{\label{Hquench} 
(color online). Time evolution of the charge density after a field quench for the nearest-neighbor spin-ice model ($J=0$). The results at $T=0.1$ and $0.01$ are plotted with $(+)$ and $(\times)$ symbols. These two temperatures lead to almost identical relaxation curves.
}
\end{figure}

As shown in Fig.~\ref{Hquench}, the charge density decreases monotonically after the field quench, within a time scale of $\mathcal{O}(1)$. The decrease of charge density is dominated by a simple pair-annihilation process of charges. This process accompanies the reduction of energy $4J_1$, and takes place within the same order of time scale.


\subsection{Field quench of the nearest-neighbour dumbbell model ($J=J_{2}=J_{3}$)}
\label{field_quench}

From now on, we shall consider the {stochastic dynamics} of Hamiltonian (\ref{chargeHamiltonian}) ($J_2=J_3\equiv J$) after a field quench of a saturated $[111]$ field: from very large $|{\mathbf H}|=100$ to ${\mathbf H}=0$. As presented previously, it is important that the initial configuration at $t=0$ is made of alternating positive and negative charges on the diamond lattice formed by the centers of the tetrahedra [see configuration in Fig.~\ref{initialconfig111}]. Thus not only is the magnetization fully saturated in the [111] direction; the charge degrees-of-freedom are also perfectly aligned. The multiple energy barriers associated with the most relevant dynamical processes are summarized in appendix \ref{dynamical_processes}.

\subsubsection{$-0.20<J<0.00$}
\label{field_quench_mJ020}
\begin{figure}[h]
\begin{center}
\includegraphics[width=0.5\textwidth]{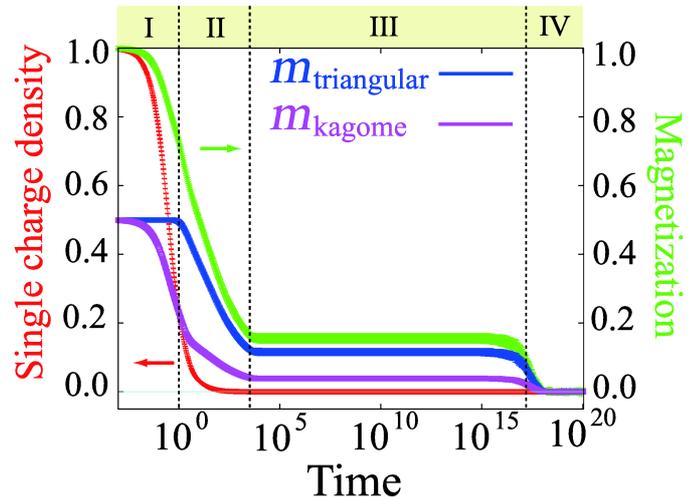}
\end{center}
\caption{\label{dynamics_mJ0.10} 
(color online). Field quench process at $J=-0.10$ and $T=0.10$. The charge density and magnetization are plotted together with the sublattice magnetizations on kagome and triangular layers. The time domain is divided into four regions I, II, III and IV, as shown with dashed lines, according to the qualitative nature of the dynamics. The sublattice magnetizations, $m_{\rm triangular}$ and $m_{\rm kagome}$, are measured along the [111] direction and are normalized by the total saturated magnetization, giving rise to the value of $0.5$ at $t=0$.
}
\end{figure}

We start with the negative region, $J<0$. The simple comparison of Figs.~\ref{Hquench} and~\ref{dynamics_mJ0.10} for $J=0$ and $J=-0.1$ respectively shows that the introduction of $J$ qualitatively alters the dynamical properties of the system. In Fig.~\ref{dynamics_mJ0.10}, the total charge density and magnetization of the system are plotted
together with the sublattice magnetization decomposed into triangular and kagome planes [see Fig.~\ref{initialconfig111} for definitions].

The charge density decays monotonically and quickly reaches practically zero on a time scale of $\mathcal{O}(1)$, in a similar way to the case of $J=0$. But in contrast to {non-interacting} spin ice, the magnetization shows a nontrivial time dependence with a wide plateau region. The dynamical process can be qualitatively divided into four time domains.

The time evolution starts with a nucleation process. Since the pyrochlore lattice is decomposed into triangular and kagome planes [Fig.~\ref{initialconfig111}], the initial event should either be a spin flip on a triangular plane or a kagome plane. The former process is unlikely to occur, since the spin flip on triangular plane gives rise to double charges $Q=\pm 4$ whose energy cost is $\Delta_5=12-12|J|>0$ [see appendix \ref{dynamical_processes} for details]. In contrast, the spin flip on the kagome plane lowers the energy, $\Delta_1=-4+20|J|<0$, which is why this process occurs within the time scale of order $\mathcal{O}(1)$.

\begin{figure*}[t]
\centering\includegraphics[width=0.75\textwidth]{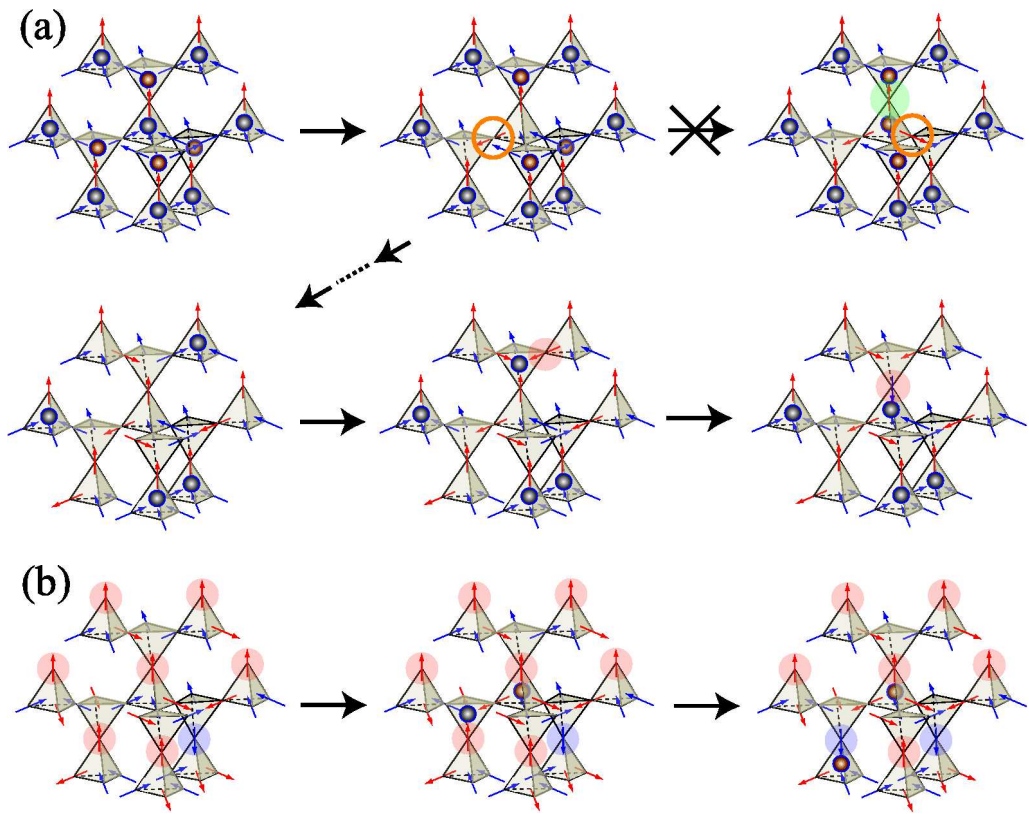}
\caption{\label{processes_mJ0.10}
(color online). Illustration of the relaxation dynamics for $-0.2<J<0.0$. (a) The initial process that occurs in the saturated configuration is the spin flip on the kagome plane, which leads to the pair annihilation of charges. While the diffusion of vacuum tetrahedra is energetically unfavorable, successive pair annihilations are possible, making the charge density more and more dilute. After enough dilution, the charges start to diffuse, as highlighted by red circles. (b) Before the magnetization relaxes to zero, charges are exhausted. The relaxation of magnetization is then only possible via pair-creation and dissociation of charges. Spin flips are highlighted by yellow circles when necessary.
}
\end{figure*}

Consequently in the time domain I of Fig.~\ref{dynamics_mJ0.10}, spins flip almost randomly but only on kagome layers, creating pairs of vacuum tetrahedra -- two-in two-out states. This is confirmed by Fig.~\ref{dynamics_mJ0.10} where only the magnetization on kagome planes decreases, while that on triangular planes remains constant. However, these spin flips are not completely random. Indeed, one could have thought that the created pairs of vacuum tetrahedra can dissociate and diffuse within their kagome planes. However, vacuum tetrahedra are confined for $J<0$ because their diffusion brings same-sign charges next to each other, a process whose energy cost is roughly proportional to the distance between the diffused pair of vacuum tetrahedra [see Fig.~\ref{processes_mJ0.10}]. This process is thus prohibited at the beginning of the relaxation (time domain I), as opposed to the $J=0$ case. 

The spin flips on kagome layers strongly suppress the density of charges in the system [see Fig.~\ref{dynamics_mJ0.10}]. Once the charges become dilute enough, they can move freely, without contact with other charges. This is the beginning of the second time domain where, in particular, charges can tunnel between adjacent kagome layers by flipping spins on triangular planes, resulting in the decrease of the triangular sublattice magnetization.

However, in addition to the diffusion of charges, their density continues to decrease, which  eventually causes an exhaustion problem. The system may not have had the time to fully decorrelate before the exhaustion of charges, leaving a long time plateau with finite magnetization and no charges. This is the third time domain of Fig.~\ref{dynamics_mJ0.10}. Please note that this time plateau may split into two parts for very small temperatures [see discussion in appendix~\ref{appendix_plateau}].

\begin{figure}[h]
\begin{center}
\includegraphics[width=0.45\textwidth]{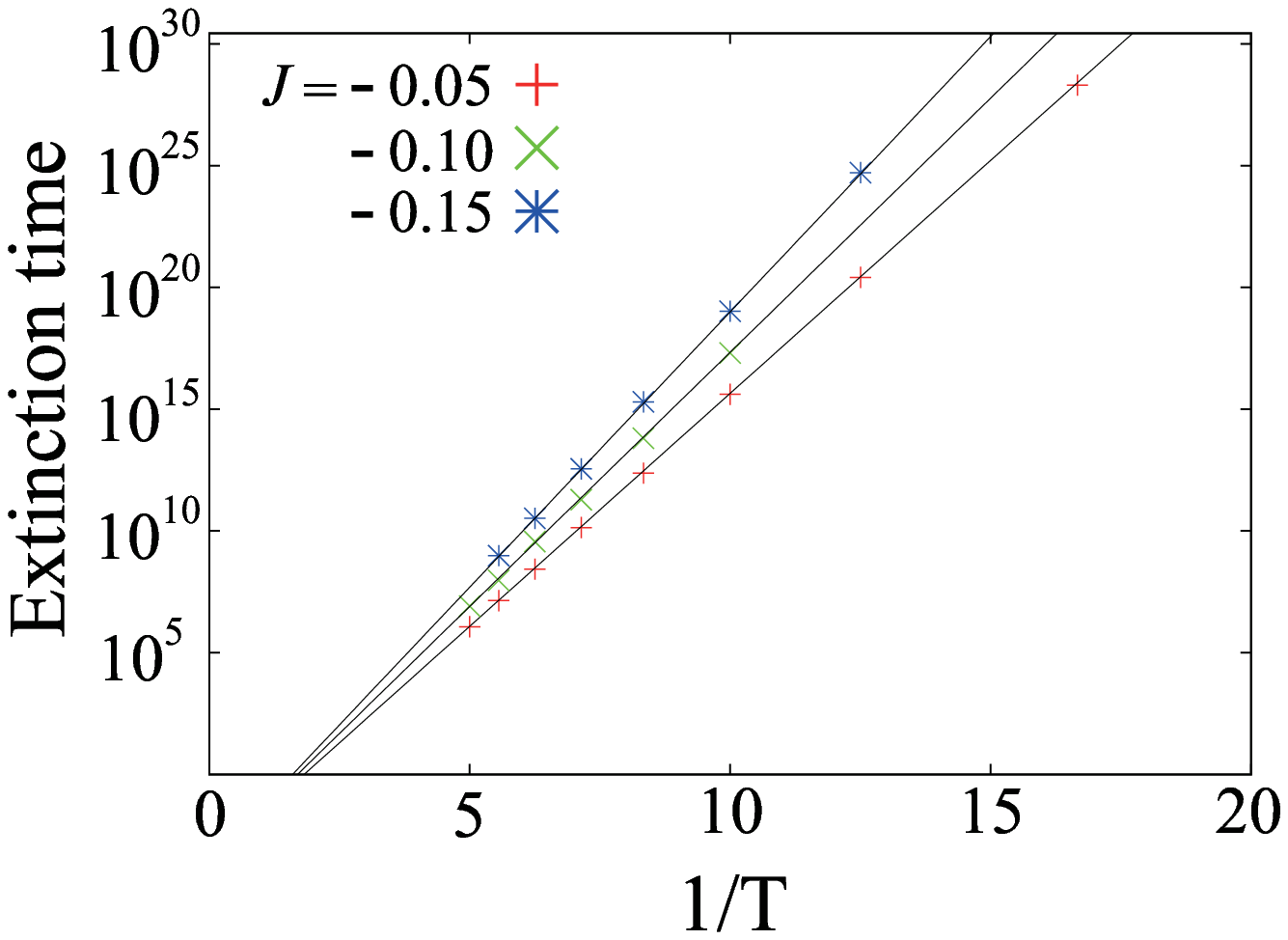}
\end{center}
\caption{\label{mJ0.10_finaldecay_scaling} 
(color online). The extinction time $t_p$ of the magnetization is plotted against the inverse of temperature $T$ for $J=-0.05, -0.10$ and $-0.15$. $t_p$ is defined as the time at which magnetization takes half of the plateau value. The $t_p$'s follow the predicted Arrhenius law, Eq.~(\ref{middlenegativeJ_Arrhenius}) with $\log t_p =  -8.00531 + 4.4/T$, $-8.12707 + 4.8/T$ and $-8.31068 + 5.2/T$
for $J=-0.05, -0.10$ and $-0.15$, respectively.
}
\end{figure}

The system then remains frozen until the creation \textit{and} dissociation of charge pairs spontaneously occur, on a time scale of the order of $\exp[(\Delta_3+\Delta_4)/T]$, where $\Delta_3=4+4|J|$ and $\Delta_4=4|J|$. The dissociated charges propagate through the system and bring the magnetization to its equilibrium vanishingly small value, defining the fourth and final time domain.  In Fig.~\ref{mJ0.10_finaldecay_scaling}, the time $t_p$ at which the charge density finally decays to zero is plotted as a function of $1/T$. $t_p$ clearly follows the predicted scaling law
\begin{eqnarray}
t_p\propto\exp[(\Delta_3+\Delta_4)/T]=\exp[(4+8|J|)/T].
\label{middlenegativeJ_Arrhenius}
\end{eqnarray}
This confirms that the lifetime of the magnetization time plateau is determined by the pair creation and dissociation process of charges.

To summarize so far, the nonequilibrium relaxation dynamics for $-0.2<J<0.0$ displays the rich multiscale behaviour, well understood from microscopic processes. However, in light of the possibility of spontaneous Hall effect where we are looking for broken time-reversal symmetry without magnetization, this region is not particularly relevant since magnetization persists on a much longer time scale than the charge density.

\begin{figure}[h]
\begin{center}
\includegraphics[width=0.45\textwidth]{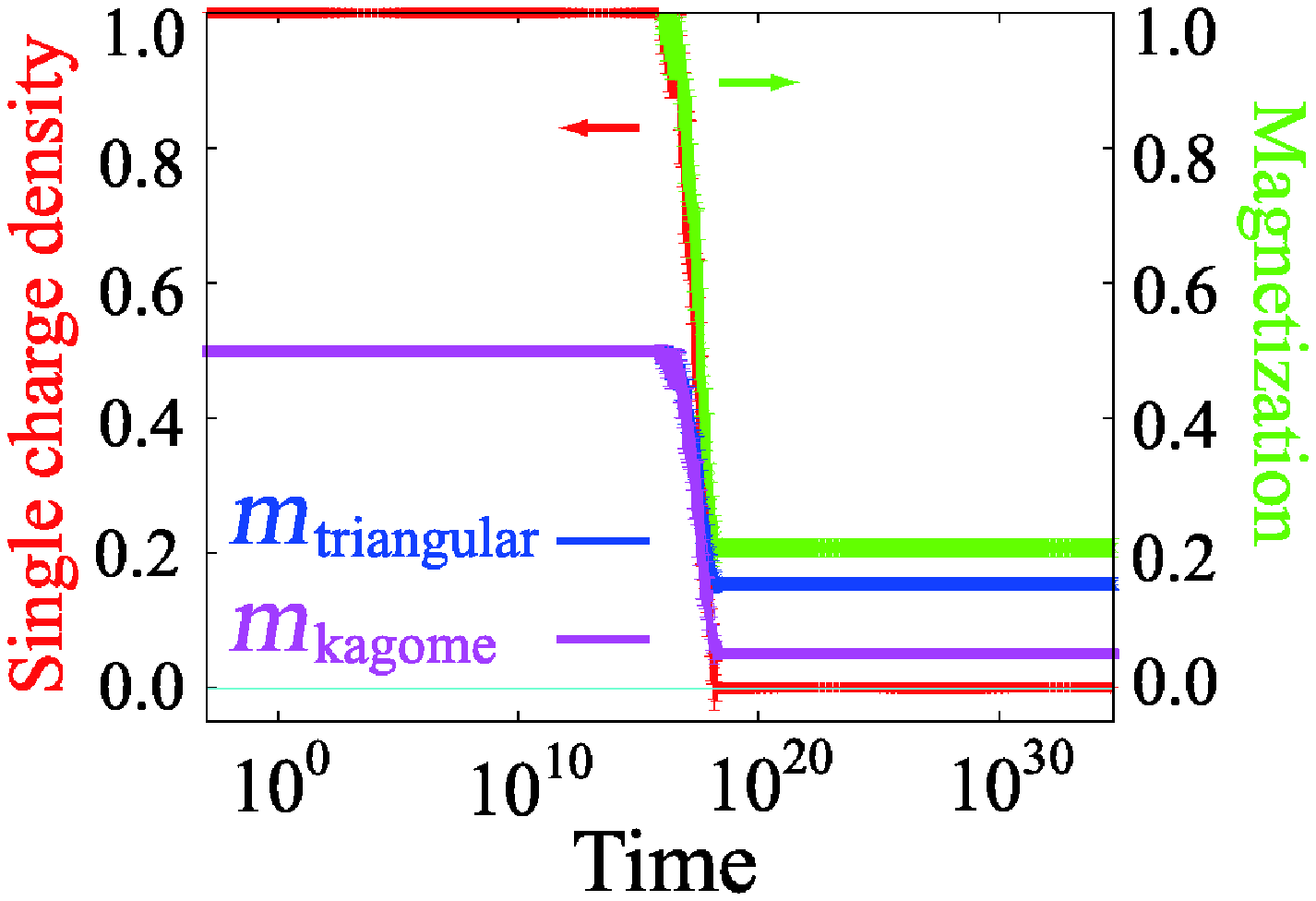}
\end{center}
\caption{\label{dynamics_mJ0.20} 
(color online). Field quench process at $J=-0.225$ and $T=0.01$. The charge density and magnetization are plotted together with the sublattice magnetizations on kagome and triangular layers. The sublattice magnetizations, $m_{\rm triangular}$ and $m_{\rm kagome}$, are measured along the [111] direction and are normalized by the total saturated magnetization, giving rise to the value of $0.5$ at $t=0$.
}
\end{figure}
\subsubsection{$-0.25<J<-0.20$}
\label{field_quench_mJ025}
The nature of dynamics qualitatively changes at $J=-0.20$. Both magnetization and single charge density remain saturated over a time scale growing exponentially with $1/T$ after the magnetic field is switched off at $t=0$ [Fig.~\ref{dynamics_mJ0.20}]. Their decay is then simultaneous and visible in both triangular and kagome layers. After this decay, a wide plateau follows and the dynamics becomes similar to the region for smaller $|J|$.

The persistence of the initial state can be attributed to the dynamical bottleneck of the initial process. As discussed in the section \ref{field_quench_mJ020}, the initial process is given by the spin flip on the kagome plane, which costs $\Delta_1=-4+20|J|$. Consequently, for $J<-0.20$, this process requires a finite energy and takes an exponentially long time at low temperatures.

However, once a pair of charges is created, the next process becomes immediately available, since it costs no more energy to create a pair of vacuum tetrahedra adjacent to the nucleation pair: $\Delta_2=-4+16|J|<0$. Consequently the first spin flip serves as a nucleation seed for an avalanche effect of charge annihilation, until the charges completely disappear from the system. After the exhaustion of charges, the system reaches a plateau with residual magnetization before finally relaxing thanks to the creation and dissociation of charges, as observed for $-0.2<J<0.0$.

\subsubsection{$-0.50<J<-0.25$}
\begin{figure*}[t]
\begin{center}
\includegraphics[width=0.75\textwidth]{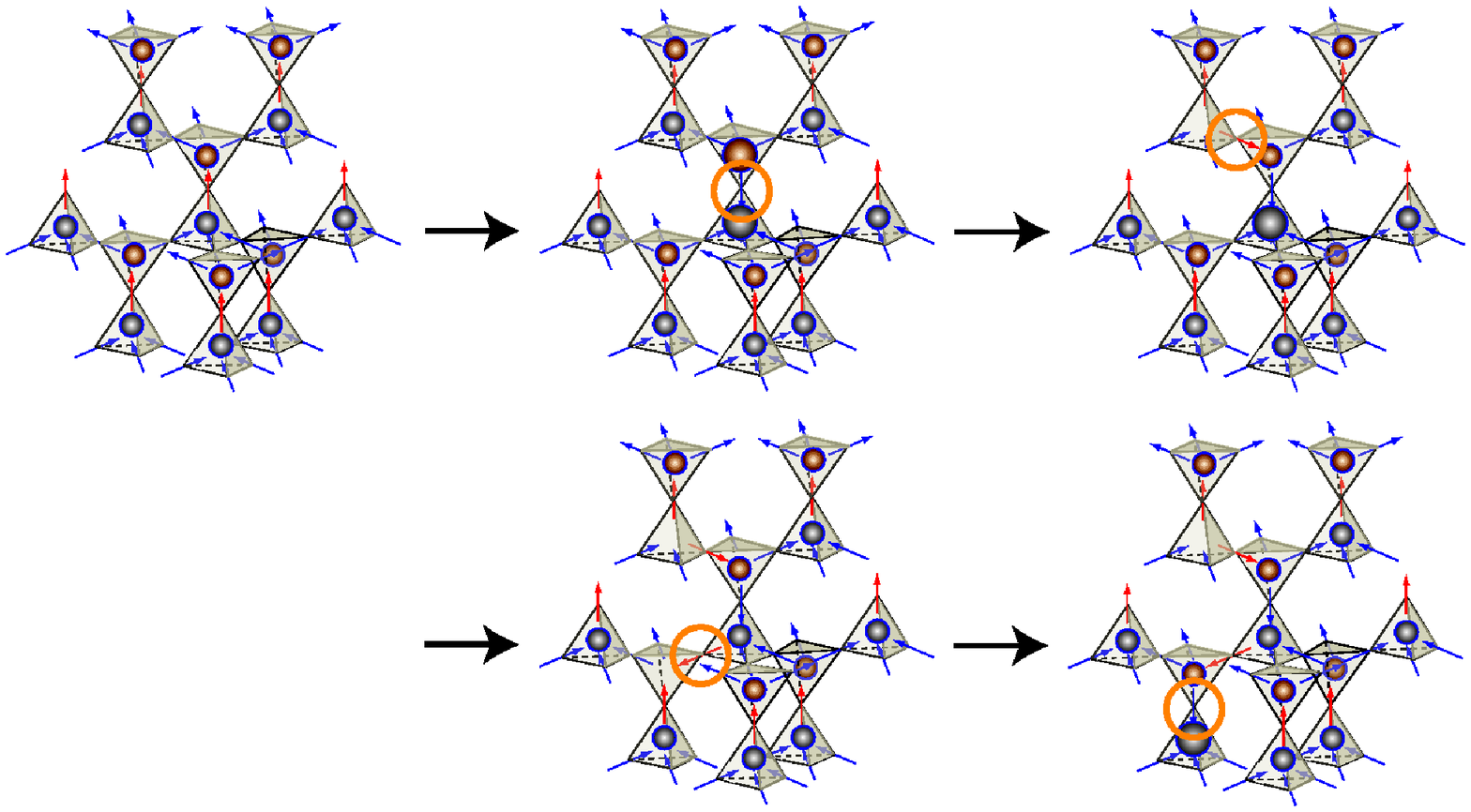}
\end{center}
\caption{\label{mJ0.45schematics} 
(color online). Illustration of the nucleation process for $-0.5<J<-3/7$ where the defects propagate in the [111] direction. It allows for spin decorrelation from the initial saturated configuration, while preserving the long-range charge order [see Fig.~\ref{monopolecrystal}].
}
\end{figure*}

\paragraph{Dynamics}
In this parameter range, both the nucleation ($\Delta_1=-4+20|J|>0$) and the proliferation ($\Delta_2=-4+16|J|>0$) processes cost a finite energy, preventing the avalanche effect that takes place for $-0.25<J$. As an alternative mechanism of relaxation, one needs to consider the dynamics along the [111] direction, perpendicular to the kagome planes. The first two steps correspond to the creation of a vacuum tetrahedron and a double charge [see Fig.~\ref{mJ0.45schematics}]. The energy of this two-step process is $\Delta_{1}+\Delta_{9}=\Delta_{5}+\Delta_{6}=4+8|J|$ and is independent of whether the nucleation occurs in a kagome or a triangular plane [see the appendix \ref{dynamical_processes} for details]. When $J\geqslant-3/7\approx -0.43$, this two-step mechanism along the [111] direction remains more expensive in energy than the two-step proliferation in kagome planes previously considered, $\Delta_{1}+\Delta_{2}\leqslant 4+8|J|$. But the dynamics after these two initial steps is mediated by coexisting energy barriers with similar magnitudes, and the system is subject to a rather complex dynamical competition for $-3/7<J<-0.25$ which varies from short to long length scales.

On the other hand for $-0.5<J<-3/7$, the mechanism along the [111] direction becomes the most favorable source of relaxation, both for their two-step nucleation ($\Delta_{1}+\Delta_{9}=\Delta_{5}+\Delta_{6} < \Delta_{1}+\Delta_{2}$) and their propagation ($-\Delta_{7}=\Delta_{8}=8-16|J| < \Delta_{2}$) [see Fig.~\ref{mJ0.45schematics}].

\begin{figure}[t]
\begin{center}
\includegraphics[width=0.45\textwidth]{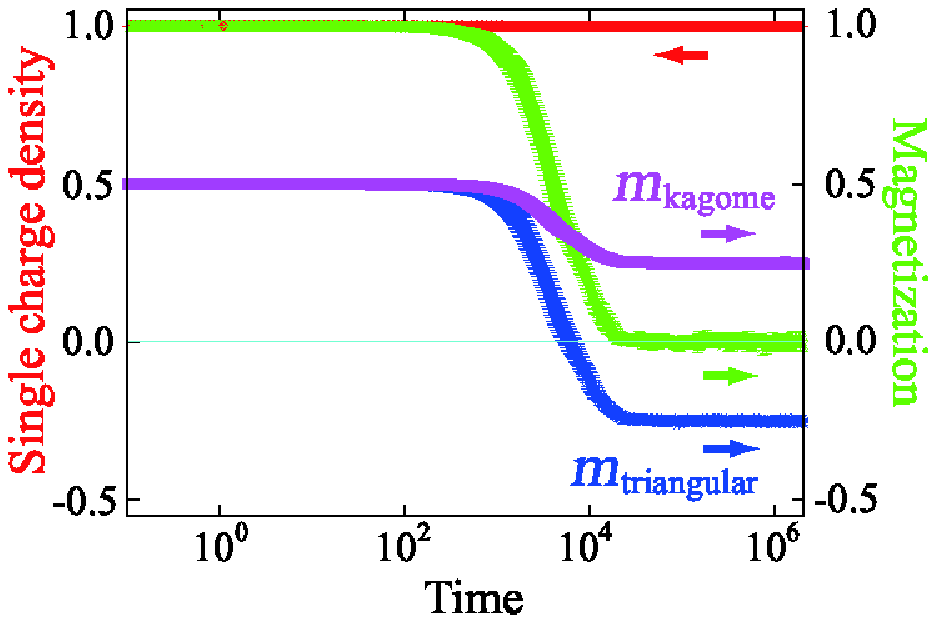}
\end{center}
\caption{\label{mJ0.45} 
(color online). Field quench process at $J=-0.45$ and $T=0.50$, showing the charge density and total magnetization as well as the sublattice magnetizations on kagome and triangular layers. The second plateau for $t\gtrsim 10^{4}$ corresponds to the metastable fragmented Coulomb spin liquid phase. The sublattice magnetizations, $m_{\rm triangular}$ and $m_{\rm kagome}$, are measured along the [111] direction and are normalized by the total saturated magnetization, giving rise to the value of $0.5$ at $t=0$.
}
\end{figure}
\begin{figure}[h]
\begin{center}
\includegraphics[width=0.45\textwidth]{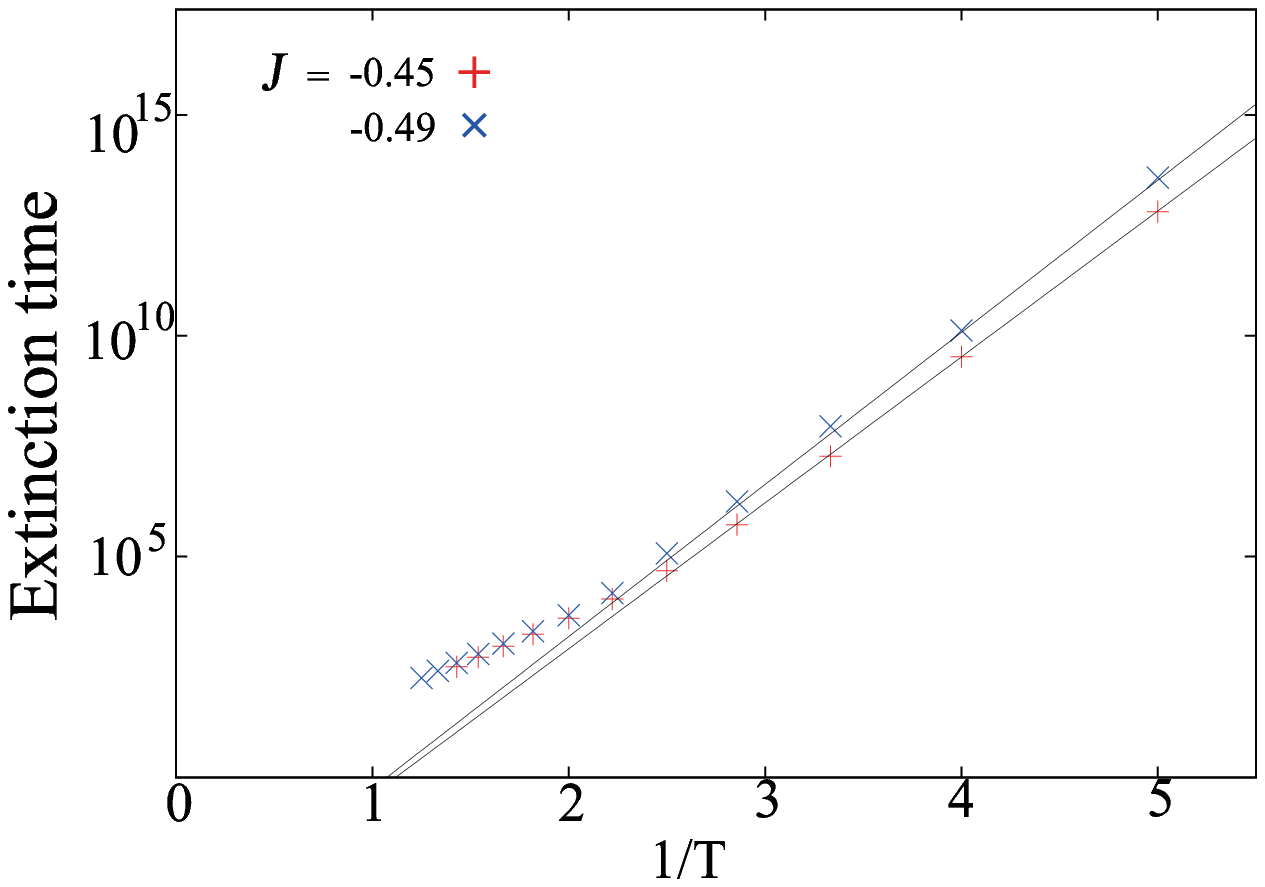}
\end{center}
\caption{\label{mJ0.49} 
(color online). The extinction time $t_p$ of the total magnetization for $J=-0.45$ and $-0.49$ as a function of the inverse temperature. $t_p$ is fitted with $\log t_p=-8.5+7.6/T$ and $-8.5+7.92/T$, respectively.
}
\end{figure}

This scenario is confirmed by simulations in Figs.~\ref{mJ0.45} and~\ref{mJ0.49}. Most appreciably, the total magnetization drops to zero, while i) the charge density keeps its saturated value for several orders of magnitude in time and ii) the sublattice magnetizations on triangular and kagome layers both take finite values. The extinction time of the total magnetization is in quantitative agreement with an Arrhenius law $\exp[-(\Delta_5+\Delta_6)/T] = \exp[-(4+8|J|)/T]$ at low temperature, reflecting the magnitude of the energy barrier of the nucleation process.

\begin{figure}[t]
\begin{center}
\includegraphics[width=0.3\textwidth]{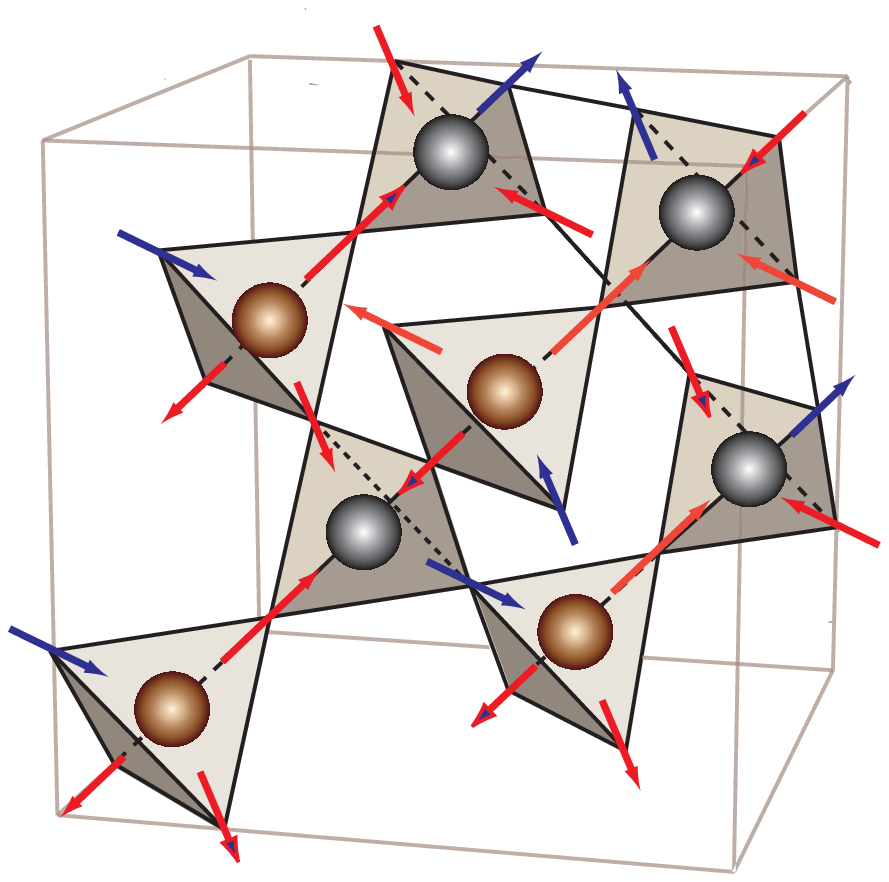}
\end{center}
\caption{\label{monopolecrystal} 
(color online). An example configuration of the fragmented Coulomb spin liquid (FCSL)~\cite{Borzi2013,Brooks2014,Jaubert2015}. All upward (downward) tetrahedra are occupied by positive (negative) charges. However, this long-range charge order does not prevent an extensive spin degeneracy with emergent Coulomb gauge theory.
}
\end{figure}

As for the plateau of Fig.~\ref{mJ0.45} after the extinction of the total magnetization, it comes from the fact that, as opposed to the proliferation of vacuum tetrahedra observed for $-0.25\leqslant J$, the mechanism along the [111] direction is mediated by the propagation of pairs of local excitations: vacuum tetrahedra and double charges~\cite{Jaubert2015}. Both excitations can fragment into two single charges: one which is fixed and respects the initial charge order, remnant of the field quench, and the second one which can move along the [111] direction. As a result, the density of single charges varies only very weakly, in order to accommodate a vanishingly small density of local excitations. For the parameter region $-0.5<J<-3/7$, the propagation along the [111] direction costs zero or little energy $-\Delta_{7}=\Delta_{8}=8-16|J|\in [0;1.14]$ when compared to the nucleation process $\Delta_{1}+\Delta_{9}=\Delta_{5}+\Delta_{6}=4+8|J|\in[7.43;8]$. As a consequence the system does not relax into the vacuum ground state made of two-in two-out tetrahedra [see Fig.~\ref{Basic}], but conserves the initial long-range charge order illustrated in Fig.~\ref{monopolecrystal}.\\

\paragraph{Fragmented Coulomb spin liquid}
The FCSL phase is partially ordered due to a broken $\mathcal{Z}_{2}$ symmetry of the charge degrees-of-freedom: every positive charge has four negative nearest-neighbour charges, and vice-versa [see Fig.~\ref{monopolecrystal}]. This leaves a $\mathcal{Z}_{2}$ degeneracy of the charge order, as for the all-in / all-out phase. But as opposed to the all-in / all-out order, the magnetic ``crystal'' of the FCSL phase is made of single charges, which retains an extensive spin degeneracy. It also supports an emergent Coulomb gauge field due to a local divergence-free constraint on the magnetization flux~\cite{Brooks2014}. When averaged over its Gibbs ensemble, the FCSL phase does not bear any finite total magnetization. However, the broken $\mathcal{Z}_{2}$ symmetry of the charge order gives rise to finite sublattice magnetizations along the [111] direction. Please note that this broken symmetry is not spontaneous, but selected by the initial [111] magnetic field. As illustrated on Fig.~\ref{initialconfig111}, a spin on a triangular layer always has a negative charge above, and a positive one below, resulting in an averaged triangular magnetization $m_{\rm triangular}=-1/4$. On the other hand, a spin on a kagome layer always has a positive charge above, and a negative one below, resulting in an averaged kagome magnetization $m_{\rm kagome}=+1/4$. These values correspond to the plateaux observed in Fig.~\ref{mJ0.45} for $t>10^{4}$, and imply that a substantial part of the FCSL Gibbs ensemble is visited during the relaxation.

A ferromagnetic analogue of the FCSL phase has been predicted in the magnetization plateau of HgCr$_{2}$O$_{4}$ and CdCr$_{2}$O$_{4}$ materials~\cite{Penc2004,Bergman2006}, where the (partial) order parameter couples to the magnetic field. However, the present FCSL phase is an antiferromagnet. 

With the peculiar exceptions of models where double charges were explicitly forbidden~\cite{Borzi2013,Brooks2014}, and 2D artificial kagome ice systems~\cite{MoellerMoessner2009, Chern2011, Arnalds12,Zhang2013,Anghinolfi15}, the FCSL has been noticeably elusive at equilibrium in 3D spin-ice models~\cite{Guruciaga2014}, despite its deceptive simplicity.

In the present nearest-neighbour dumbbell model for example, it is part of the ground-state ensemble only for $J=-0.5$, degenerate with the all-in/all-out order and the two-in two-out spin-ice ground state [see Fig.~\ref{Basic}]. It disappears however at finite temperature in favor of the two-in two-out spin-ice regime because of their entropy difference. The present setup thus offers a rare mechanism able to stabilize a metastable form of the FCSL phase, expanding the versatility of nonequilibrium physics in spin ice~\cite{Mostame2014}.\\

\paragraph{Hall response}
The final question regarding the FCSL phase is its relevance to the Hall effect. For this parameter range, the relaxation of the charge density is much slower than that of the magnetization [see Fig.~\ref{mJ0.45}]. In this sense, the situation may be similar to the experimental condition of the spontaneous Hall effect observed in Pr$_{2}$Ir$_{2}$O$_{7}$~[\onlinecite{Machida10}]. However, based on symmetry arguments, the Hall conductivity is exactly zero in the FCSL phase. Quite generally, provided the spin configurations on upward and downward tetrahedra can be transformed into each other through time-reversal, then the Hall conductivity becomes zero, if spatial homogeneity is preserved on average [see appendix~\ref{appendix_Hall}]. This means that a finite Hall response is not expected to occur for negative $J$, where positive charges tend to sit next to negative ones.

\subsubsection{$0.00<J<0.25$}
\begin{figure}[b]
\begin{center}
\includegraphics[width=0.45\textwidth]{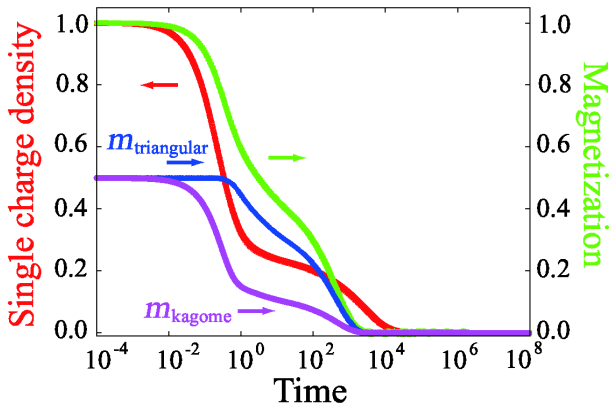}
\end{center}
\caption{\label{positiveJ} 
(color online). The relaxation of the charge density and magnetization after the field quench are plotted for several combinations of $J$ and $T$, keeping the ratio $J/T=1.25$ fixed: $(J, T)=(0.05, 0.04), (0.10, 0.08), (0.15, 0.12)$ and $(0.20, 0.16)$. All relaxation data quantitatively collapse onto a single curve, suggesting a scaling relation with a single parameter, $J/T$.
}
\end{figure}

Now, let us consider the other side of the phase diagram, with $J>0$. Charges of opposite sign repel, making the initial $[111]$ polarized state particularly unstable. The dynamical feature after the field quench is plotted in Fig.~\ref{positiveJ}, where the decrease in charge density $\rho_1$ and total magnetization $M$ start very quickly. After a time scale of $\mathcal{O}(1)$, both $\rho_1$ and $M$ develop a shoulder-like feature after which the relaxation process slows down. Nevertheless, the magnetization decays smoothly to zero, and subsequently, charges also disappear completely. Interestingly, the whole time dependence of $\rho_1$ and $M$ can be scaled with a single parameter $J/T$ [see Fig. \ref{positiveJ}].

The early stage of the relaxation is very similar to the $-0.20\leqslant J\leqslant 0$ case of section~\ref{field_quench_mJ020}. There is no energy barrier preventing the relaxation from the initial $[111]$ polarized state. Vacuum tetrahedra quickly proliferate in kagome layers, diminishing the kagome magnetization in the process. Once the density of charges becomes small enough, spins in the triangular layers can finally relax without creating pairs of double charges. However, as opposed to the $J<0$ scenario, nearest-neighbour pairs of positive/negative single charges can separate at no energy cost since they repel. There is thus no energy barrier for the diffusion of charges. The resulting separation of charges makes their eventual annihilation statistically more scarce and energetically unfavorable since a pair of positive/negative charges needs to get close to each other before being able to annihilate. Accordingly, in the absence of high-energy charge creation processes, the dynamics in this time domain is dominated by a single energy scale, the nearest-neighbour repulsion, proportional to $J$ [see equation~(\ref{chargeHamiltonian})], which explains the $J/T$ scaling and the quasi-plateau of charge density observed in Fig. \ref{positiveJ}. Furthermore, even if charge diffusion first slows down the magnetization decay (the diffusion process is mostly uncorrelated with the direction of the spin), it ultimately allows for the complete relaxation of the system down to zero magnetization.\\

\begin{figure}[t]
\begin{center}
\includegraphics[width=0.45\textwidth]{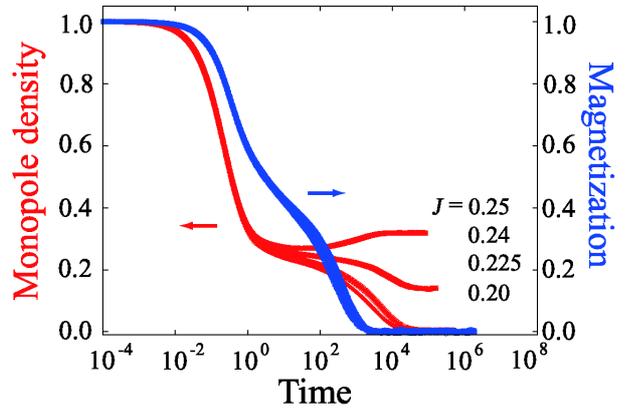}
\end{center}
\caption{\label{break_scaling} 
(color online). The charge density and magnetization are plotted for several combinations of $J$ and $T$, keeping the ratio $J/T=1.25$ fixed: $(J, T)=(0.25, 0.20), (0.24, 0.192), (0.225, 0.18)$ and $(0.20, 0.16)$. Deviation of charge density from the scaling curve is observed for $J\sim0.25$, implying the breakdown of the one parameter scaling relation and the appearance of charge creation process.
}
\end{figure}

For larger values of $J$, the magnetization shows approximately the same curve as in Fig.~\ref{positiveJ}, but the previous $J/T$ scaling breaks down for the charge density $\rho$ on long time scales [see Fig.~\ref{break_scaling}]. It implies the appearance of a new energy barrier which is not simply proportional to $J$, \textit{i.e.} a dynamical mechanism involving charge creation/annihilation [see Eq.~(\ref{chargeHamiltonian})]. This is understood by the fact that the creation cost of charges in equation~(\ref{chargeHamiltonian}) diminishes as $J$ increases. The creation of a pair of charges can be further facilitated if it occurs in the appropriate ``cage'' of same-sign charges. This is why, for large enough $J$, the energy barrier of the nearest-neighbour repulsion which was responsible for the $J/T$ scaling can be replaced by a charge-creation barrier. Finally, when $J$ reaches the value of $1/4$, the relaxation of $\rho_1$ becomes non-monotonic, underlining a qualitative change of physics at $J=1/4$ as discussed in section~\ref{section_phase_diagram}.

\section{Jellyfish point at $J=1/4$}
\label{thermodynamics}

\subsection{High symmetry point of the Hamiltonian}
\label{sec:highsym}

\begin{figure}[t]
\begin{center}
\includegraphics[width=0.45\textwidth]{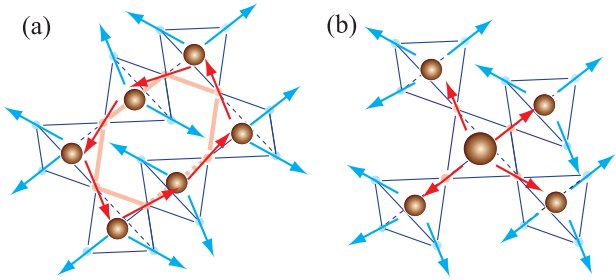}
\end{center}
\caption{\label{MinimalCluster} 
(color online). Schematic figures of the minimal clusters: (a) a ring made of six tetrahedra for the jellyfish and (b) methane-like structure.}
\end{figure}

At $J=1/4$, the nearest-neighbour dumbbell model becomes
\begin{eqnarray}
\mathcal{H} = \frac{1}{4}\sum_{p}Q_p^2 - \frac{1}{4}\sum_{\langle p,q\rangle}Q_p Q_q.
\label{chargeHamiltonian_critical}
\end{eqnarray}
Hence, the energy cost for creating a charge is exactly balanced with the energy gain from placing a pair of same-sign charges next to each other. As briefly introduced in section~\ref{section_phase_diagram}, there are two minimal clusters of spins respecting this balance. The first one is a closed ring of charges, where the number of charges is equal to the number of bonds between them. The ring is made of at least six tetrahedra. The second type of cluster is reminiscent of a methane molecule, composed of a double charge next to four single charges [see Fig.~\ref{MinimalCluster}].

These two clusters are actually the seeds of extended zero-energy textures. Attaching same-sign charges to these seeds does not cost energy since it adds the same number of charges and bonds. As illustrated in Fig.~\ref{schematic_jellyfish}, it is possible to attach branches to the central ring, which we pictorially refer to as a \textit{jellyfish} structure. Consequently, for $J=1/4$, the two-in two-out spin-ice ground state receives additional macroscopic degeneracy coming from these composite forms. Away from the $J=1/4$ point, these structures become excitations with energy:
\begin{align}
&E_{\rm jellyfish} = 4(N_{\rm branch} + 6)\Bigl(\frac{1}{2} - 2J\Bigr),
\label{eq:Ejelly}\\
&E_{\rm methane} = 4(N_{\rm branch} + 8)\Bigl(\frac{1}{2} - 2J\Bigr),
\end{align}
where $N_{\rm branch}$ is the number of charges in the branch.

If the charge description is elegant, one should not forget the underlying spin texture.
As illustrated in Fig.~\ref{schematic_jellyfish}, each tetrahedron in a branch hosts a gauge charge, and therefore has three majority spins and one minority spin. Each tetrahedron is connected to its ``parent'' tetrahedron strictly and uniquely via the minority spin, which means that the branches can bifurcate or trifurcate, but never form closed loops nor connect to other jellyfish or methane-like structures. Such connection is energetically unfavorable for opposite-sign structures and topologically forbidden for same-sign structures. Moreover, the sequence of spins along the backbone of any branch form an oriented path. Depending on the sign of the gauge charges, this oriented path flows either towards or away from the central seed (the ring or the double charge).

For the jellyfish structure, there is an additional degree-of-freedom present in the central ring, which carries a circular magnetization flow with two choices of chirality: clockwise or counterclockwise. Due to this circular magnetization flow, the ring acquires a toroidal moment, ${\mathbf T}_{ring}$, defined by
\begin{eqnarray}
{\mathbf T}_{ring} = \sum_i {\mathbf r}_i\times{\mathbf S}_i,
\end{eqnarray}
where, ${\mathbf r}_i$ is the relative coordinate of site $i$, measured from the center of the ring. Consequently, ${\mathbf T}$ flips its sign when the magnetization flow is reversed. It is an important property which allows the \textit{local} encoding of time-reversal symmetry in a zero-energy texture, independently of the sign of the gauge charges it belongs to, i.e.~independently of global time-reversal symmetry.

This emergent toroidal moment takes an enhanced flavor in the context of field-quench dynamics discussed in the previous section. If a coupling between spin and spatial degrees of freedom is taken into account -- e.g.~by adding spin-orbit coupling in our model -- the field induced time-reversal symmetry breaking of the initial state can be potentially encoded in the ring of the jellyfish structures. In other words, toroidal moments may persist long after the magnetic field is quenched. It is tempting to associate this possibility with the spontaneous Hall effect observed in Pr$_2$Ir$_2$O$_7$ [\onlinecite{Machida10}]. Namely the residual Hall response may be ascribed to the combined effects of slow relaxation and toroidal ordering due to the jellyfish clusters.
Indeed, the decay of a jellyfish cluster incurs an activation energy barrier of the order $J$ and becomes exponentially slow in $J/T$ at low temperatures (see Appendix~\ref{collapse_of_jellyfish} for a discussion of the possible decay processes and their relative barriers). 
\\

\begin{figure}[t]
\begin{center}
\includegraphics[width=0.45\textwidth]{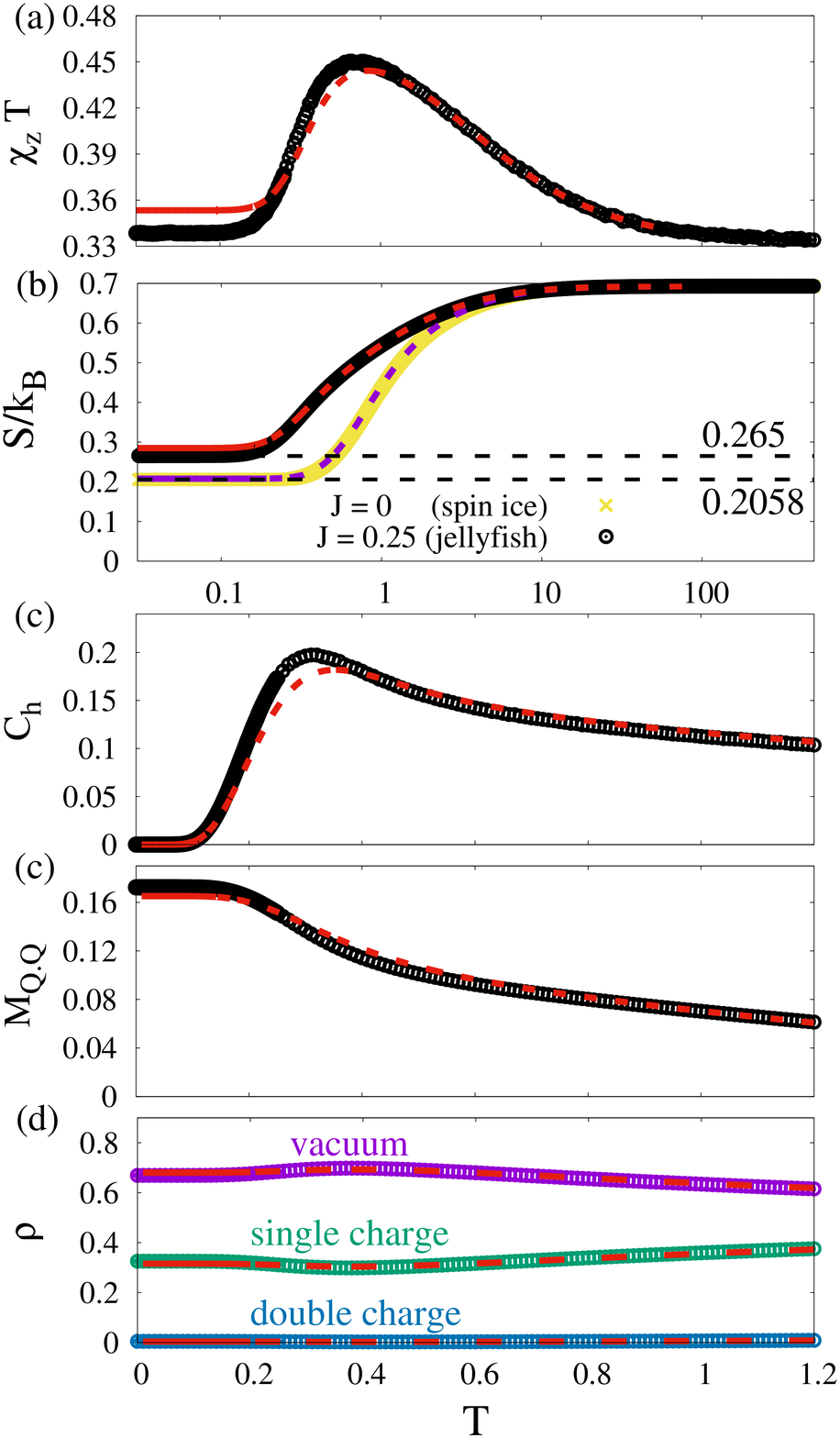}
\end{center}
\caption{\label{fig:obsJ0.25} 
(color online). Absence of long-range order for $J=1/4$, probed by Monte Carlo simulations (black circles) and Bethe lattice calculations (red dashed lines). (a) Magnetic susceptibility along the $z-$axis times temperature. We observe a Curie law crossover without any singularities (\textit{i.e.} without any phase transition), characteristic of a classical spin liquid at low temperature~\cite{Jaubert2013}. (b) Entropy of the jellyfish model compared to the nearest-neighbor spin-ice model (yellow crosses). The black dashed lines show the low-temperature limits from Monte Carlo and confirm the higher entropy of the jellyfish ground state. (c) Specific heat. (d) Quasi-order parameter defined in eq.~\ref{eq:MQQ}, showing short-range ordering with a finite fraction of same-sign nearest-neighbour monopoles. (e) Density of tetrahedra configurations. All observables are normalized per number of spins, except for panel (e) which is normalized per number of tetrahedra. The temperature axis is on a logarithmic scale for panels (a,b).}
\end{figure}
\subsection{Crossover into the jellyfish regime}
\label{sec:cross}

We have studied the emergence of the jellyfish structures by the means of Monte Carlo simulations and Bethe-lattice approximation at equilibrium. Technical details of the methods are discussed in appendices~\ref{appendix_MC} and~\ref{appendix_Bethe}. 

Let us first define the variable
\begin{eqnarray}
M_{Q.Q}=\frac{1}{N}\sum_{\langle p,q \rangle} \frac{Q_{p}.Q_{q}}{4}.
\label{eq:MQQ}
\end{eqnarray}
where $N$ is the total number of pyrochlore sites. $M_{Q.Q}$ is a spatially averaged measurement of the nearest-neighbour ordering of charges. In the all-in/all-out antiferromagnetic order and in the FCSL phase, it takes the values $M_{Q.Q}=-4$ and $-1$ respectively. Please note that $M_{Q.Q}$ is not a proper order parameter of these phases, since {$M_{Q.Q}$ does not} differentiate between the configurations with broken $\mathcal{Z}_{2}$ symmetry. But it is a convenient probe of the evolution of nearest-neighbour correlation and it directly couples to the interaction term of equation~(\ref{chargeHamiltonian_critical}).

$M_{Q.Q}$ takes a finite positive value as expected for $J=1/4>0$, and reaches a plateau for $T<0.2$ [see Fig.~\ref{fig:obsJ0.25}.(d)]. This plateau is a signature of the jellyfish regime and it is also revealed by a corresponding plateau in the charge density $\rho$ in Fig.~\ref{fig:obsJ0.25}.(e), as confirmed by the quantitative agreement between Monte Carlo simulations and Bethe-lattice calculations [see the dashed lines in Fig.~\ref{fig:obsJ0.25} for the Bethe-lattice results].

The jellyfish regime is characterized by the following observables as $T\rightarrow 0^{+}$:
\begin{eqnarray}
&\rho_{0}&=0.670\pm 0.005,\\
&\rho_{1}&=0.324\pm 0.005,\\
&\rho_{2}&=0.005\pm 0.001,\\
&M_{Q.Q}&=0.171\pm 0.005,\label{eq:J14T0M}\\
&\chi_{z}\,T&=0.333\pm 0.003.
\end{eqnarray}
Let us check if these values are consistent with our understanding of section~\ref{sec:highsym}. Let $N$ and $N_{t}$ be the number of spins and tetrahedra respectively, with $N=2 N_{t}$. Then there are $N_{t}\rho_{2}$ double charges, each of them surrounded by four single charges [see section~\ref{sec:highsym}]. Thus $4 N_{t}\rho_{2}$ single charges are directly linked to a double charge, while $N_{t}(\rho_{1}-4 \rho_{2})$ are linked to a single charge in the branches of jellyfish or methane structures. One can thus estimate
\begin{eqnarray}
M_{Q.Q}^{\rm est}=\dfrac{N_{t}\left[\left(\rho_{1}-4 \rho_{2}\right)+2\left(4 \rho_{2}\right)\right]}{N}\nonumber=\dfrac{\rho_{1}+4 \rho_{2}}{2}\approx 0.172.
\label{eq:MQQest}
\end{eqnarray}
which quantitatively matches the measured value of equation~(\ref{eq:J14T0M}). Please note that the above formula is also valid for Husimi calculations, but with somewhat lower values of $M_{Q.Q}=0.165$, $\rho_1=0.316$ and $\rho_2=0.00363$.

For $J=1/4$, the ground-state ensemble does not break any symmetry of the Hamiltonian, and includes all of the two-in two-out configurations (traditional spin-ice ground states) and all of the configurations with the structures discussed in section~\ref{sec:highsym} (jellyfish and methane-like). The entropy of the $J=1/4$ ground state is thus extensively large and higher than the one of spin ice [Fig.~\ref{fig:obsJ0.25}.(b)]. The absence of symmetry breaking and the very high entropy justify why the passage from the high-temperature paramagnet to the jellyfish regime is a crossover and not a phase transition, as confirmed by the specific heat $C_{h}$ and the magnetic susceptibility $\chi_{z}$ of Fig.~\ref{fig:obsJ0.25}.(a,c).

The variance of the magnetization $\langle \Delta M_{z}^{2}\rangle\equiv\langle M_{z}^{2}\rangle-\langle M_{z} \rangle^{2}\equiv N\chi_{z}\,T$ directly measures the build-up of correlations in classical spin liquids, visible in the Curie-law crossover of Fig.~\ref{fig:obsJ0.25}.(a)~\cite{Jaubert2013}. The fact that it reaches the value of 1/3 (within error bars) as $T\rightarrow0^{+}$, \textit{i.e.} as for a standard paramagnet, is remarkable and consistent with the fact that the long-range dipolar correlations of the two-in two-out phase~\cite{Isakov2004} are cut off by the jellyfish and methane-like structures.

In the absence of a critical point, the Bethe lattice calculations provide a very good approximation of the real system. This is especially visible in Fig.~\ref{fig:obsJ0.25}.(e), where the temperature dependence of the charge density is essentially indistinguishable from simulations. There is however a small but clear difference for the second-order cumulant of the energy and magnetisation [see ~\ref{fig:obsJ0.25}.(a,c)]. This is probably a consequence of the \textit{de facto} absence of closed loops of spins in the Bethe lattice, preventing the emergence of jellyfish structures.

\begin{figure}[h]
\begin{center}
\includegraphics[width=0.5\textwidth]{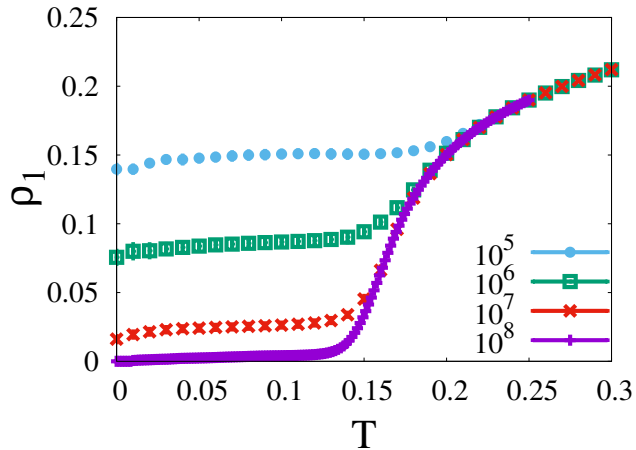}
\end{center}
\caption{\label{fig:obsJ0.24} 
(color online). Temperature dependence of the density of single charges probed by Monte Carlo simulations for $J=0.24$, just below the jellyfish phase. Each curve corresponds to a different simulation time, from $10^{5}$ to $10^{8}$ Monte Carlo steps. This shows that even if the thermalization process is very slow, the density of charges eventually vanishes for $J<1/4$. We use the definition of a Monte Carlo step, made of $N_t$ single-spin flip attempts and 10 worm updates, with $N_t$, the number of tetrahedra. Monte Carlo dynamics include 1 parallel tempering every 100 Monte Carlo steps.
}
\end{figure}
\begin{figure}[t]
\begin{center}
\includegraphics[width=0.49\textwidth]{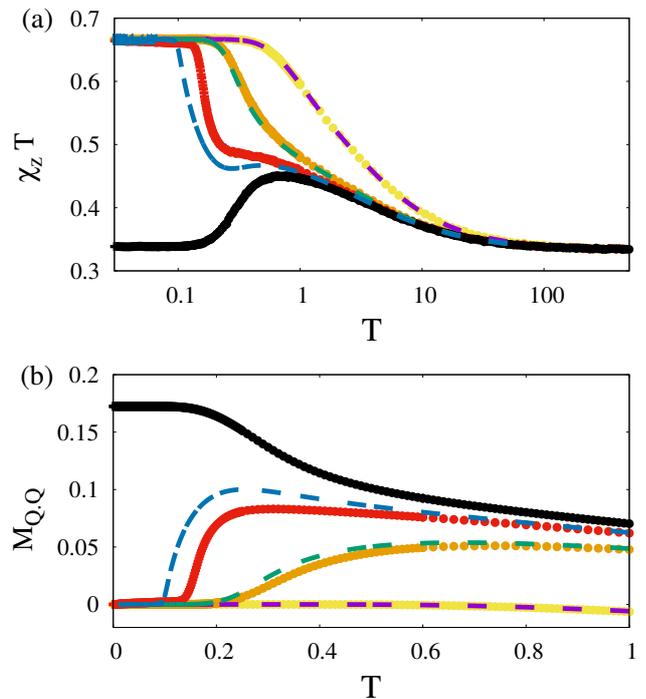}
\end{center}
\caption{\label{fig:obsJX} 
(color online). Temperature dependence of the thermodynamics probed by Monte Carlo simulations for $J\in\{0,0.22,0.24,0.25\}$ (yellow, orange, red and black respectively). The dashed lines are analytical Husimi calculations. (a) Magnetic susceptibility along the $z-$axis times temperature. The horizontal temperature axis is on a logarithmic scale. We observe two different kinds of Curie law crossover. For $J<1/4$, $\chi_{z}T$ ultimately goes to $\approx 2/3$, characteristic of the two-in two-out Coulomb phase~\cite{Jaubert2013}. For $J=1/4$, $\chi_{z}T$ shows a clear downturn towards $\approx 1/3$, signature of a classical spin liquid different from the Coulomb phase. (b) Order parameter $M_{Q.Q}$ defined in eq.~\ref{eq:MQQ}, showing partial ordering for $J=1/4$, but not $J<1/4$. As $J$ increases, $M_{Q.Q}$ remains strictly positive to lower and lower temperature, showing the growing influence of the jellyfish phase. The agreement between analytics and simulations is quantitative for $J=\{0,0.22\}$ and semi-quantitative for $J=0.24$, where the influence of jellyfish structures at finite temperature comes into play -- especially the ring of single charges.
}
\end{figure}
\begin{figure}[t]
\begin{center}
\includegraphics[width=0.49\textwidth]{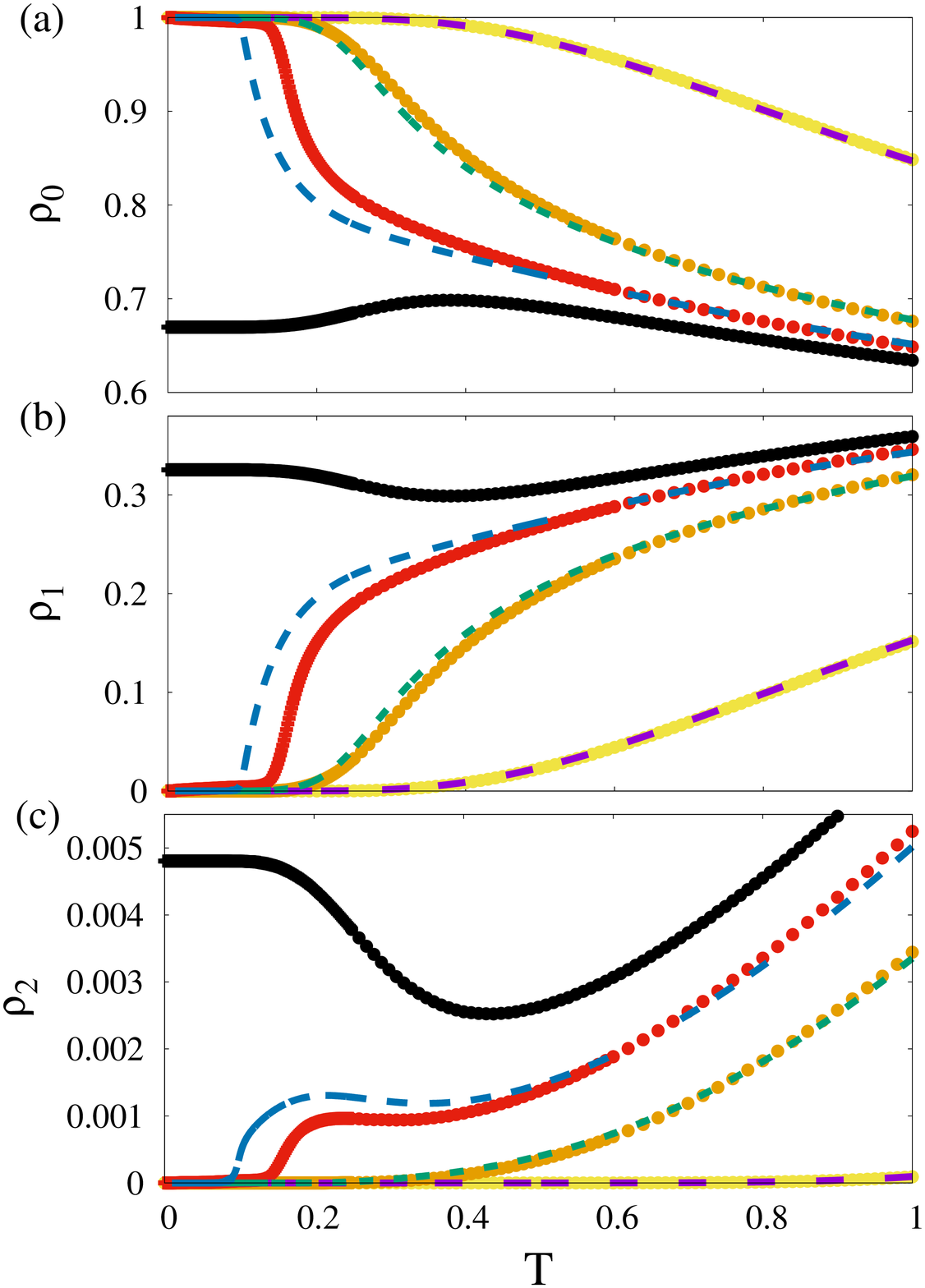}
\end{center}
\caption{\label{fig:rhoJX} 
(color online). Temperature dependence of the densities of vacuum tetrahedra (a), single (b) and double (c) charges probed by Monte Carlo simulations for $J\in\{0,0.22,0.24,0.25\}$ (yellow, orange, red and black respectively) and Husimi calculations (dashed lines).}
\end{figure}
\subsection{On the way to the jellyfish: $J \lesssim 1/4$}
\label{sec:ontheway}

The increase of single-charge density in Fig.~\ref{break_scaling} can now be understood by the emergence of charge structures precisely at $J=1/4$. However, even though these structures are excitations for $J<1/4$, they visibly play a role in the relaxation just below $1/4$ [see the $J=0.24$ curve in Fig.~\ref{break_scaling}]. We have confirmed this influence by Monte Carlo simulations in Fig.~\ref{fig:obsJ0.24}, where the low-temperature single-charge density $\rho_{1}$ tends to zero with increasing Monte Carlo time. Despite the use of parallel tempering and worm algorithm [see appendix~\ref{appendix_MC}], and a very slow decrease of the temperature during the preliminary Monte Carlo equilibration -- made of up to 10 million Monte-Carlo steps (MCs) -- simulations require 100 million MCs to reach the equilibrium vanishing value of $\rho_{1}$. Such slow dynamics in absence of disorder is well-known in spin ice~\cite{Matsuhira2001,Snyder2001,Jaubert2009,Mostame2014,Slobinsky2010,Castelnovo2010,Levis2012,Levis2013}, even without any thermal or field quench, but relatively rare for a phase with a finite density of charges, which are usually the source of dynamics [see \textit{e.g.} the third time domain of Fig.~\ref{dynamics_mJ0.10}].

Both single and double charges completely vanish as $T\rightarrow 0$ for $0\leqslant J<1/4$ [see Fig.~\ref{fig:rhoJX}], but with a pronounced shoulder at finite temperature as one gets closer to the $1/4$ value. This shoulder marks the change from the low-temperature two-in two-out regime, and the entropically induced apparition of jellyfish structures at intermediate temperatures. The crossover temperature is characterized by the energy scale $E_{\rm jellyfish}$ of Eq.~(\ref{eq:Ejelly}). This behaviour is most visible in the Curie-law crossover of Fig.~\ref{fig:obsJX}.(a). The quantity $\chi_{z}\,T$ always reaches the $2/3$ value characteristic of spin ice~\cite{Jaubert2013}, but delayed by an order of magnitude in temperature between $J=0$ (yellow curve) and $J=0.24$ (red curve). The same abrupt crossover is visible in $M_{Q.Q}$ [see Fig.~\ref{fig:obsJX}.(b)]. The presence of a crossover instead of a phase transition is confirmed by Bethe-lattice calculations, but could not be completely ruled out by simulations for $0.24 < J < 0.25$. On a side note, the small negative value of $M_{Q.Q}$ for $J=0$ (nearest-neighbor spin-ice model) is a consequence of the entropic Coulomb interactions between topological defects in spin ice~\cite{Henley2010,Castelnovo2011,Jaubert2015}.

\begin{figure*}[t!]
\centering\includegraphics[width=18cm]{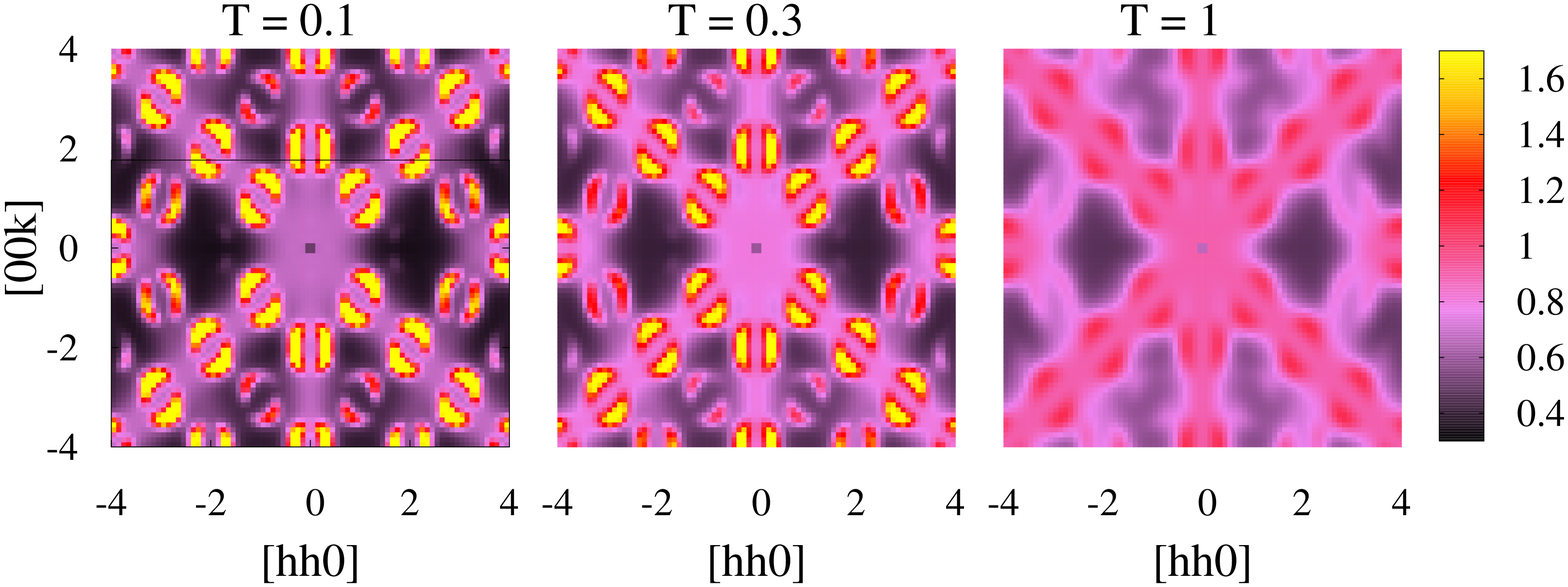}\\
\centering\includegraphics[width=18cm]{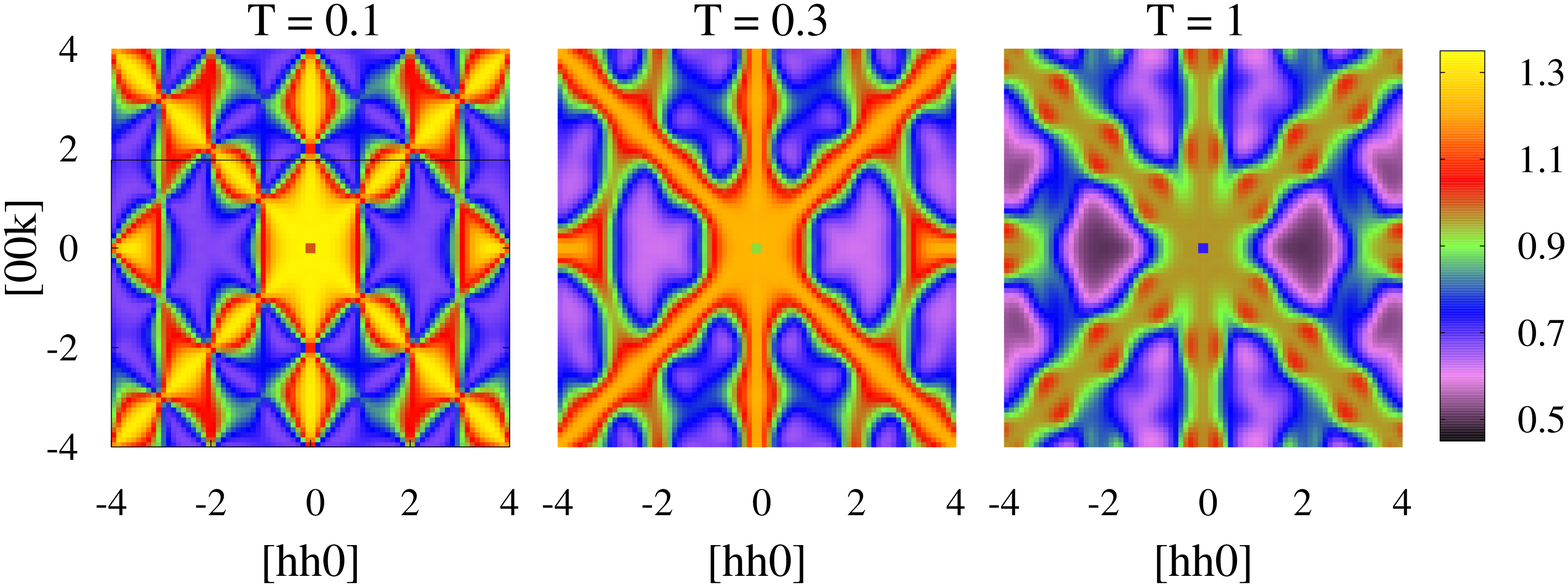}
\caption{\label{fig:SQMC} 
(color online). Structure factor $S(\mathbf{q})$ in the [hhk] plane as measured by neutron scattering obtained from Monte Carlo simulations for $J=0.25$ (\textit{top}) and $0.22$ (\textit{bottom}) at $T=\{0.1,0.3,1\}$. \textit{Top}: For $J=1/4$, the pattern of the structure factor is relatively similar from $T=1$ to $0.1$, showing sharper intensities around the half-moon features upon cooling (in yellow at $T=0.1$). \textit{Bottom}: For $J=0.22$ on the other hand, one moves continuously from the low-temperature spin-ice regime with sharp pinch points~\cite{Fennell2009} to the jellyfish regime at $T=1$ [see section~\ref{sec:ontheway}]. The half-moon features characteristic of the jellyfish regime sit at the same wavevectors as the pinch points in spin ice. The color scale is the same for all panels for each given value of $J$.
}
\end{figure*}
\begin{figure}[t]
\centering\includegraphics[width=\columnwidth]{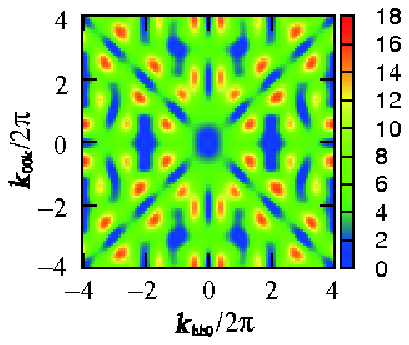}
\caption{\label{MonopoleRingImage} 
(color online). Contribution from the charge rings to the spin flip part of the structure factor. We assume that all rings are made of six tetrahedra and that the four possible orientations of hexagonal rings are equally populated [see appendix \ref{ring_to_SofQ}]. {Remnants of ``half-moon" structure can be clearly seen around the Brillouin zone centers, consistent with the results of Monte Carlo simulation [Fig.~\ref{fig:SQMC}]}.
}
\end{figure}
\subsection{Neutron-scattering signature}

Our final question concerns the experimental detection of the jellyfish regime. A possible fingerprint would be the Curie-law crossover of Fig.~\ref{fig:obsJ0.25}.(a). But since $\chi_{z}\,T$ reaches a value of $1/3$ identical to a standard paramagnet, the real signature lies in the bump of the crossover which unfortunately could be modified by perturbations pertinent at intermediate or high temperatures such as finite single-ion anisotropy.

On the other hand, quasi-elastic neutron scattering should be able to detect the fingerprints of the jellyfish structures. Neutron scattering measures the structure factor of the material
\begin{eqnarray}
S(\mathbf{q})=\frac{1}{N}\left|\sum_{i} \mathbf{S}_{i}^{\perp}\, e^{i \mathbf{q}\cdot\mathbf{r}_{i}}\right|^{2},
\label{eq:SQ}
\end{eqnarray}
where $\mathbf{S}_{i}^{\perp}$ are the spin components orthogonal to the vector $\mathbf{q}$ and $\mathbf{r}_{i}$ is the position of site $i$. The structure factor obtained from neutron scattering for $J=1/4$ is shown in Fig.~\ref{fig:SQMC} (top panels). At low temperature, distinctive half-moon features of scattering appear. These features are relatively broad and surround -- \textit{i.e.} do not sit on -- the Brillouin zone centers. They persist at higher temperatures, albeit more diffuse, and are confirmed by a phenomenological analysis taking into account only the contributions from the ring part of the jellyfish [see Fig.~\ref{MonopoleRingImage}]. These half-moon scattering motifs are very exotic for a spin-ice model and can thus serve as a solid signature of the jellyfish regime in experiments.

When stepping away from the $J=1/4$ model, one recovers the characteristic pinch points of the spin-ice Coulomb phase at very low temperature [see Fig.~\ref{fig:SQMC} (bottom panels) for $J=0.22$], as predicted from the discussion of section~\ref{sec:ontheway}. At higher temperatures ($T=1$), the higher entropy of the jellyfish regime wins over the Coulomb phase, and one recovers the half-moon features of $J=1/4$. In particular this broad scattering precisely replaces the pinch points, consistent with the correlated density of charges in the jellyfish regime~\cite{Fennell2009,Sen2013}.

\section{Conclusion}
\label{conclusion}

Motivated by the coupling between itinerant electrons and localized spins in spin-ice materials, we have investigated the physics of Ising spins on the pyrochlore lattice with first, second- and third-neighbour interactions [see equation~(\ref{spinHamiltonian})]. We have focused on the line where $J_2=J_3=J$, where our model can be rewritten in terms of charge degrees-of-freedom, giving rise to the nearest-neighbour dumbbell model [see equation~(\ref{chargeHamiltonian})].

The two-in two-out Coulomb phase remains the ground state of the model over a broad range of parameters ($-0.5<J<0.25$). By varying $J$ within this window, the relaxation processes after a field quench in the [111] direction display a rich diversity of nonequilibrium phenomena, with \textit{e.g.} glassy behavior in absence of disorder due to multi-scale energy barriers, which can be quantitatively understood by a microscopic approach.\\

For small values of $J$ ($-0.25<J<0$), the magnetization decay shows markedly slow relaxation when compared to the charge density. It gives rise to (multiple) magnetization plateaux over several orders of magnitude in time due to the exhaustion of charges. When approaching from $J=-0.25$ to the phase-diagram boundary $J=-1/2$, the dynamics changes drastically and is dominated by the diffusion of double charges and vacuum tetrahedra. As a result, the initial charge order imposed by the field quench at $t=0$ persists while the magnetization quickly vanishes. The resulting metastable phase is known as the fragmented Coulomb spin liquid (FCSL)~\cite{Borzi2013,Brooks2014,Jaubert2015}, where long-range (charge) order co-exists with a (Coulomb) spin liquid. The FCSL is a very elusive phase, which makes the present field-quench protocol a rare opportunity to realize it in a realistic model.
 
 As for positive $J$, the dynamics develop an interesting $J/T$ scaling law due to the repulsion between opposite-sign charges, which hinders their ability to annihilate pairwise. As approaching $J=1/4$, this scaling breaks down as the system acquires a finite gauge charge density in thermodynamic equilibrium, due to the high population of composite excitations which we have pictorially termed methane and jellyfish, as discussed in Sec.~\ref{sec:highsym}. These structures, once created, decay slowly because of the high energy barriers accompanied with their collapse [see Sec.~\ref{collapse_of_jellyfish}].
 
While the jellyfish structure is non-magnetic, the central ring carries an emergent chiral degree of freedom, characterized by a toroidal moment. Accordingly, the jellyfish structure is able to encode the information of time-reversal symmetry breaking with this toroidal degree of freedom. This feature, combined with the long lifetime, implies a potential relevance to the spontaneous Hall effect observed in Pr$_2$Ir$_2$O$_7$.
 
 Aiming at the detection of jellyfish, we combined numerical/phenomenological analyses to show that the ``half-moon" structure in the $S({\mathbf q})$ is a clear signature of the jellyfish structure. Experimentally, the detection of various types of composite excitations has been carried out for frustrated magnets, such as Zn/Mg/HgCr$_2$O$_4$ and Tb$_2$Ti$_2$O$_7$\cite{SHLee2000,SHLee2002,Yasui2002,Tomiyasu2008,Tomiyasu2011,Tomiyasu2013}. Our analysis indicates that the quasi-elastic neutron scattering is a promising experimental probe to detect the jellyfish, which may clarify the origin of spontaneous Hall effect in Pr$_2$Ir$_2$O$_7$.

The natural next step of this analysis would be to investigate what happens for $J>1/4$. In this region the two-in two-out Coulomb phase is not part of the ground state anymore, and the ordering mechanism is completely dominated by the attraction between same-sign charges, inducing strong steric and kinetic constraints. Our simulations indicate an increase in single and double charges below a first order transition, but we found that this phase is very difficult to thermalize.

Beyond the original motivation of RKKY interactions in metallic systems, our model directly applies to insulating pyrochlores with farther-neighbour exchange. Such interactions might be present in spin-ice materials~\cite{Yavorskii2008,McClarty2015,Henelius2016} and be of importance for quantum spin ices where virtual crystal field excitations can induce coupling beyond nearest-neighbors~\cite{Molavian2007,Chou2010}. The $J_{2}=J_{3}$ condition is an arguably strong constraint, but very useful by its simplicity in order to serve as a basis for perturbative approaches.

Our paper also provides a working example of what happens when it is possible to tune the interactions between topological defects away from their natural setting. Indeed, spin ice has often been described as a magnetic analogue of an electrolyte~\cite{Castelnovo2010b,Zhou2011,Giblin2011,Castelnovo2011,Kaiser2015}, a picture which relies on the Coulomb attraction between opposite-sign charges. The possibility to reverse the sign of this interaction, even just at the nearest-neighbour level, drastically changes the physics of the problem. Clusters of quasi-particles can be stabilized because i) charges of same signs cannot annihilate by pair and ii) charges of opposite signs -- which could annihilate -- are repelled from each other. Such questions are certainly not limited to spin ice, but extend naturally to phases prone to emergent quasi-particles such as quantum spin liquids, topological phases and artificial gauge fields.

Last but not least, the apparition of the FCSL phase demonstrates the promises of engineering macroscopic states via nonequilibrium techniques~\cite{Paulsen2014}. We hope our work will further motivate the exploration of nonequilibrium phenomena in geometrically frustrated magnets, ranging from topological ergodicity breaking~\cite{Castelnovo2007} to glassiness without disorder~\cite{Jaubert2009,Castelnovo2010}, and with connections to kagome systems~\cite{Cepas2012,Cepas2014} including their realization in artificial spin ice~\cite{MoellerMoessner2009,Chern2011,Arnalds12,Zhang2013,Anghinolfi15} and the 16-vertex model~\cite{Levis2012,Levis2013}.

\begin{acknowledgments}
It is a pleasure to thank John Chalker for useful discussions. This work was supported by DFG via SFB 1143, Engineering and
Physical Sciences Research Council (EPSRC) Grant No. EP/G049394/1 (C.C.), the Helmholtz Virtual Institute ``New
States of Matter and Their Excitations," the EPSRC NetworkPlus on “Emergence and Physics far from Equilibrium,”
the Okinawa Institute of Science and Technology Graduate University and by JSPS KAKENHI (Nos. 26400339, 24340076, 15H05852 and 15K13533).
\end{acknowledgments}

\appendix

\section{Charge representation of the $J_1-J_2-J_3$ spin ice model}
 \label{derive_charge}
\begin{figure}[h]
\begin{center}
\includegraphics[width=0.45\textwidth]{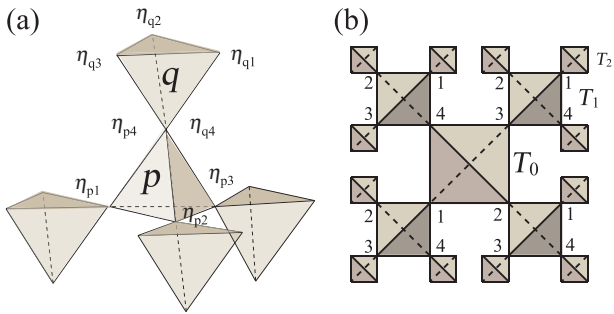}
\end{center}
\caption{\label{Fig_lattice} 
(color online). (a) The tetrahedron cluster and the convention of site indices used for the derivation of nearest-neighbour dumbbell model. (b) The tetrahedron Husimi cactus used for the Bethe approximation.
}
\end{figure}

Here, we derive the charge representation of the $J_1-J_2-J_3$ spin ice model at $J_2=J_3=J$.
For this purpose, we rewrite the Hamiltonian (\ref{spinHamiltonian}), up to constant, in the following form:
\begin{align}
\mathcal{H} = \frac{1}{2}\sum_p &(\eta_{p1} + \eta_{p2} + \eta_{p3} + \eta_{p4})^2\nonumber\\
+ J\sum_{\langle p,q\rangle}&(\eta_{p1} + \eta_{p2} + \eta_{p3} + \eta_{p4} - \eta_{[p,q]})\nonumber\\
&\hspace{0.2cm}\times(\eta_{q1} + \eta_{q2} + \eta_{q3} + \eta_{q4} - \eta_{[p,q]}).
\label{selfenergy}
\end{align}
In the first term, the summation is taken over the tetrahedra, $p$, while in the second term, the summation is over neighboring pairs of tetrahedra, $\langle p,q\rangle$.
The first term comes from the nearest-neighbor interactions which connect spins on the same tetrahedron.
Meanwhile, the second term is due to the second- and third-neighbor couplings which connect spins on neighboring tetrahedra. 
For this rewriting, we named the sites on the neighboring tetrahedra, $p$ and $q$, as $p1-p4$ and $q1-q4$, as shown in Fig.~\ref{Fig_lattice} (a).
These tetrahedra share one site, which we call $\eta_{[p,q]}$.
In the convention shown in Fig.~\ref{Fig_lattice} (a), $\eta_{[p,q]} = \eta_{p4} = \eta_{q4}$.

By introducing a charge variable, $\tilde{Q}_p\equiv\eta_{p1} + \eta_{p2} + \eta_{p3} + \eta_{p4}$,
the first term can be transformed into
\begin{eqnarray}
\frac{1}{2}\sum_p (\eta_{p1} + \eta_{p2} + \eta_{p3} + \eta_{p4})^2 = \frac{1}{2}\sum_p\tilde{Q}_p^2.
\end{eqnarray}
The second term can be written as 
\begin{align}
&J\sum_{\langle p,q\rangle}(\eta_{p1} + \eta_{p2} + \eta_{p3} + \eta_{p4} - \eta_{[p,q]})\times\nonumber\\
&\qquad\quad(\eta_{q1} + \eta_{q2} + \eta_{q3} + \eta_{q4} - \eta_{[p,q]})\nonumber\\
&=J\sum_{\langle p,q\rangle}(\tilde{Q}_p - \eta_{[p,q]})(\tilde{Q}_q - \eta_{[p,q]})\nonumber\\
&=J\sum_{\langle p,q\rangle}(\tilde{Q}_p\tilde{Q}_q - \tilde{Q}_p\eta_{[p,q]} - \tilde{Q}_q\eta_{[p,q]}) + Const.\nonumber\\
&=J\sum_{\langle p,q\rangle}\tilde{Q}_p\tilde{Q}_q -J\sum_{p}\tilde{Q}_p\sum_{q={\rm n.n.\ of}\ p}\eta_{[p,q]} + Const.\nonumber\\
&=J\sum_{\langle p,q\rangle}\tilde{Q}_p\tilde{Q}_q -J\sum_{p}\tilde{Q}_p^2 + Const.
\label{inter_tetrahedron}
\end{align}

The equations (\ref{selfenergy}) and (\ref{inter_tetrahedron}) add up to the representation,
\begin{eqnarray}
\mathcal{H} = \bigl(\frac{1}{2} - J\bigr)\sum_p\tilde{Q}_p^2 + J\sum_{\langle p,q\rangle}\tilde{Q}_p\tilde{Q}_q.
\end{eqnarray}
Finally, by introducing a new charge variable:
\begin{eqnarray}
Q_p = \tilde{Q}_p, Q_q = -\tilde{Q}_q, 
\end{eqnarray}
for upward (downward) tetrahedra, $p$ ($q$). We end up with the form,
\begin{eqnarray}
\mathcal{H} = \bigl(\frac{1}{2} - J\bigr)\sum_pQ_p^2 - J\sum_{\langle p,q\rangle}Q_pQ_q.
\end{eqnarray}


\section{Waiting time Monte Carlo method}
\label{appendix_WTMC}

Since the $J_1-J_2-J_3$ spin ice model is composed of classical degrees of freedom, 
the rule of time evolution is not determined a priori. Here, we assume a stochastic dynamics defined as follows.
Suppose that the spin configuration $\Omega_{\alpha}\equiv\{{\mathbf S_i}\}$ is realized at time $t$, with probability $P(\Omega_{\alpha})$.
$P(\Omega_{\alpha})$ evolves with time by the following stochastic equation:
\begin{align}
\frac{d}{dt}P(\Omega_{\alpha}) = \frac{1}{\tau_0}\sum_{\alpha\not=\beta}[P&(\Omega_{\beta})W(\Omega_{\beta}\rightarrow\Omega_{\alpha})\nonumber\\
 &- P(\Omega_{\alpha})W(\Omega_{\alpha}\rightarrow\Omega_{\beta})],
\label{stochastic_equation}
\end{align}
where $\tau_0$ gives the unit of time, and $W(\Omega_{\alpha}\rightarrow\Omega_{\beta})$ is the transition rate from the state $\Omega_{\alpha}$ to $\Omega_{\beta}$. Here, we assume only a single spin flip process, i.e., $W(\Omega_{\alpha}\rightarrow\Omega_{\beta})$ is finite, if and only if the state $\Omega_{\beta}$ can be obtained from $\Omega_{\alpha}$ by flipping a single spin. For the two configurations, $\Omega_{\alpha}$ and $\Omega_{\beta}$, we assume a transition rate of thermal bath type,
\begin{eqnarray}
W(\Omega_{\alpha}\to\Omega_{\beta}) = \frac{\exp(-\beta E(\Omega_\beta))}{\exp(-\beta E(\Omega_\alpha)) + \exp(-\beta E(\Omega_\beta))}.
\label{transition_rate}
\end{eqnarray}

In order to solve the stochastic equation (\ref{stochastic_equation}), we resort to the waiting time Monte Carlo method.
The procedure of this numerics is divided into several steps.
Suppose, the system is at $n$-th step with time, $t_n$, then the calculation goes as follows:
(i)\ Make a table of transition probabilities: $p_i$ for all possible Ising variable, $\eta_i$, to flip, according to the transition rate (\ref{transition_rate}).
(ii)\ determine which $\eta_i$ to flip, according to the probability $p_i/p_{\rm tot}$\ ($p_{\rm tot}=\sum_ip_i$), by generating a random number $r_1\in[0, 1]$.
(iii)\ obtain the time $\tau$ taken to make the flip by $\tau = -\frac{1}{p_{\rm tot}}\log(r_2)$\ ($r_2\in[0,1]$).
(iv)\ update the time $t_{n+1} = t_n + \tau$.
(v)\ repeat (i)-(iv) at sufficient times and make a sample average.

In the actual calculations, we typical choose the system size: $N_{\rm site}=16\times16\times16\times4$, number of steps: $N_{\rm step}=1000 N_{\rm site}$, and the number of samples: $N_{\rm sample}=100$.

\section{Monte Carlo simulations}
\label{appendix_MC}

For the characterization of the $J=1/4$ model discussed in section~\ref{thermodynamics}, we made extensive use of classical Monte Carlo simulations with single spin flip dynamics, parallel tempering and worm algorithm. The worm algorithm especially is a powerful method to decorrelate systems where single spin flip dynamics become inefficient because of a local ice rule~\cite{Barkema1998}. In spin ice, this is the case for the two-in two-out Coulomb phase. In the present study however, we are especially interested in the presence of topological defects which break this ice rule. To circumvent the problem, we used an extension of the worm algorithm developed in  Ref.~[\onlinecite{Brooks2014}].

The worm is made of a unidirectional closed chain of spins where all spins point in the same direction along the chain. It is constructed spin by spin until it closes on itself. Let us arbitrarily choose that during its construction, the worm enters a tetrahedron via a spin pointing inward. Then, the worm may enter
\begin{itemize}
\item a two-in two-out tetrahedron: there are two equivalent outward spins to exit the tetrahedron, chosen with probability 1/2.
\item a three-in one-out tetrahedron: there is one outward spin to exit the tetrahedron, chosen with probability 1.
\item a one-in three-out tetrahedron: there are three equivalent outward spins to exit the tetrahedron, chosen with probability 1/3.
\item an all-in tetrahedron: there are no possibilities to exit the tetrahedron. The worm is stopped and erased.
\end{itemize}
The procedure is repeated until the worm closes. Because such a worm is made of as many inward spins as outward spins for every tetrahedron it encounters, flipping all spins in the worm does not modify the position nor the sign of the charges. Since the Hamiltonian of our model can be written in terms of charge degrees-of-freedom only [se equation~(\ref{chargeHamiltonian})], flipping a worm does not modify the energy of the system, and detailed balance is thus naturally respected.

\section{Bethe approximation}
\label{appendix_Bethe}

Thermodynamic properties of the $J_1-J_2-J_3$ model can be semi-analytically evaluated by Bethe approximation.
This method is equivalent to approximating the pyrochlore lattice with its loopless variant, the tetrahedron Husimi cactus (THC)
as shown in Fig.~\ref{Fig_lattice} (b). Themodynamic quantities, such as specific heat and magnetic susceptibility, as well as
charge density can be evaluated exactly on this network, and its temperature dependence rather precisely matches 
the result of Monte Carlo simulation [see section~\ref{thermodynamics}].

For the sake of explanation, we here summarize our convention. Firstly, we introduce four sublattices and assign anisotropy axis, ${\mathbf d}_{\alpha} (\alpha=1-4)$, as defined in section \ref{J1J2J3spinicemodel}, to discuss magnetic susceptibility in this scheme. The sublattice index of site $i$ is denoted as $\alpha_i$.
To discuss the tetrahedron-based quantities, we choose one tetrahedron, $T_0$, and consider the array of tetrahedra, starting from $T_0$.
In particular, we define $T_N$ as the $N$-th tetrahedron from $T_0$ [Fig.~\ref{Fig_lattice}(b)].
Each tetrahedron $T_j$ has $2^4$ possible states $\zeta_j (=1-16)$ corresponding to the values of spins, $\eta$, on its four sites.
We assume that the state $\zeta$ has charge $Q[\zeta]$, and the spin at sublattice $\alpha$ is denoted as $\eta^{\alpha}[\zeta]$.

\subsection{Partition function on half cactus}
\begin{figure}[h]
\begin{center}
\includegraphics[width=0.45\textwidth]{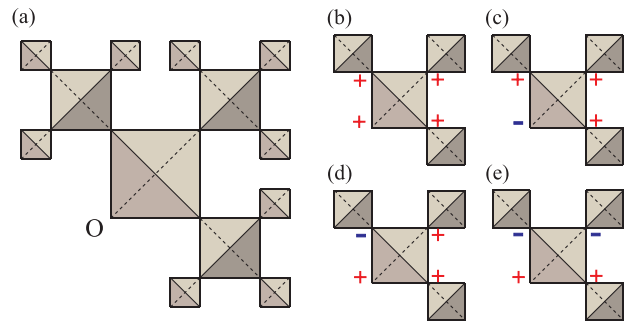}
\end{center}
\caption{\label{half_cacti} 
(color online). (a) A schematic figure of half cactus, obtained by terminating the tetrahedron Husimi cactus at site $O$. (b)-(e) The configurations of $\eta$'s on the root tetrahedron, corresponding to the partial partition function, (b) $z_4$, (c) $z_{3A}$, (d) $z_{3B}$ and (e) $z_2$. $\pm$ signs correspond to $\eta=\pm1$.
}
\end{figure}

The basic building blocks of this approximation are the partial partition functions evaluated on half cactus, which are obtained by
terminating the tetrahedron Husimi cactus at the ``root site", $O$, as shown in Fig.~\ref{half_cacti} (a).
In particular, we call the tetrahedron involving the site $O$, ``root tetrahedron".
We consider the Hamiltonian (\ref{spinHamiltonian}) on this network, and sum up the Boltzmann factors for all the configurations of $\{\eta_i\}$, to obtain the partition function, $z$.
In particular, we focus on the partial summation of Boltzmann factors, with a fixed configuration of $\eta$'s on the root tetrahedron.

As shown in Fig.~\ref{half_cacti} (b), we define $z_4$ as the partial sum of Boltzmann factors, provided all $\eta$'s have the same sign on the root tetrahedron.
Similarly, we define $z_{3A}$, $z_{3B}$ and $z_2$ as the partial summation with the 3-in 1-out or 2-in 2-out configurations on the root tetrahedron.
We distinguish $z_{3A}$ and $z_{3B}$: For the former, the root site has a minority spin, while not for the latter.
Taking account of the degeneracy of configurations on the root tetrahedron, we obtain
\begin{eqnarray}
z = 2z_4 + 6z_{3B} + 6z_2 + 2z_{3A}.
\end{eqnarray}

The remarkable feature of the half cactus is its self-similarity. By removing the root tetrahedron, one can generate 
three separate copies of the original half cactus network. By using this feature, one can prove that the partial partition functions satisfy the following recursive relations:
\begin{eqnarray}
\left\{\begin{array}{ll}
z_4 = e^{-16\beta(\frac{1}{2}-J)}\bigl[e^{-16\beta J}z_4 + 3e^{-8\beta J}z_{3B} + 3z_2 + e^{8\beta J}z_{3A}\bigr]^3,\\
z_{3B} = e^{-4\beta(\frac{1}{2}-J)}\bigl[e^{-8\beta J}z_4 + 3e^{-4\beta J}z_{3B} + 3z_2 + e^{4\beta J}z_{3A}\bigr]^2\\
\hspace{1.5cm}\times\bigl[e^{8\beta J}z_4 + 3e^{4\beta J}z_{3B} + 3z_2 + e^{-4\beta J}z_{3A}\bigr],\\
z_2 = \bigl[z_4 + 3z_{3B} + 3z_2 + z_{3A}\bigr]^3,\\
z_{3A} = e^{-4\beta(\frac{1}{2}-J)}\bigl[e^{-8\beta J}z_4 + 3e^{-4\beta J}z_{3B} + 3z_2 + e^{4\beta J}z_{3A}\bigr]^3.
\end{array}\right.
\label{selfconsistent}
\end{eqnarray}

By solving these equations, we can get three ratios, $z_4/z_2$, $z_{3B}/z_2$ and $z_{3A}/z_2$, which 
will serve as basic building blocks to obtain thermodynamic quantities, as we discuss below.

\subsection{Occupation rate}
Next, we consider the probability that a particular spin configuration is realized in a single tetrahedron.
The probability that double charge, single charge, and vacuum state is realized in a certain tetrahedron,
is proportional to $p_{2}$, $p_{1}$, and $p_{0}$, respectively, which are given by
\begin{align}
&p_2 = e^{-16\beta(\frac{1}{2}-J)}\bigl[e^{-16\beta J}z_4 + 3e^{-8\beta J}z_{3B} + 3z_2 + e^{8\beta J}z_{3A}\bigr]^4,\nonumber\\
&p_1 = e^{-4\beta(\frac{1}{2}-J)}\bigl[e^{-8\beta J}z_4 + 3e^{-4\beta J}z_{3B} + 3z_2 + e^{4\beta J}z_{3A}\bigr]^3,\nonumber\\
&\hspace{2cm}\times\bigl[e^{8\beta J}z_4 + 3e^{4\beta J}z_{3B} + 3z_2 + e^{-4\beta J}z_{3A}\bigr]\nonumber\\
&p_0 = \bigl[z_4 + 3z_{3B} + 3z_2 + z_{3A}\bigr]^4,
\end{align}
in which $z_4, z_{3B}, z_2$ and $z_{3A}$ are the solution of the self-consistent equation (\ref{selfconsistent}).

By considering the number of degeneracy, the probabilities for each spin configuration are given by
\begin{eqnarray}
\rho_{2}=\frac{2p_2}{\mathcal{N}},\ \ \rho_{1}=\frac{8p_1}{\mathcal{N}},\ \ \rho_{0}=\frac{6p_0}{\mathcal{N}}, 
\end{eqnarray}
with $\mathcal{N} = 2(p_2 + 4p_1 + 3p_0)$.

\subsection{Correlation function}
In order to evaluate physical quantities, such as specific heat and magnetic susceptibility, one needs spin or charge correlation function.
To obtain these quantities, we need the conditional probability, $P_N(\zeta'|\zeta)$: the probability that the tetrahedron $T_N$ takes the state $\zeta'$, given that $T_0$ is in the state $\zeta$.

With $P_N(\zeta'|\zeta)$, one can write down the spin correlation between site $i$ and $j$:
\begin{eqnarray}
\langle\eta_{i}\eta_{j}\rangle = \sum_{\zeta,\zeta'}P_N(\zeta'|\zeta)\eta^{\alpha_i}[\zeta]\eta^{\alpha_j}[\zeta'].
\label{spincorrelation}
\end{eqnarray}
Here, we assumed the site $i$ ($j$) belongs to the sublattice $\alpha_i$ ($\alpha_j$) of tetrahedron $T_0$ ($T_N$).

And similarly, the charge correlation between the tetrahedra, $T_0$ and $T_N$:
\begin{eqnarray}
\langle Q_{T_0}Q_{T_N}\rangle = \sum_{\zeta,\zeta'}P_N(\zeta'|\zeta)Q[\zeta]Q[\zeta'],
\label{chargecorrelation}
\end{eqnarray}

To obtain $P_N(\zeta'|\zeta)$, one can resort to the method of transfer matrix.
Suppose a $16\times16$ matrix, $K$, whose $(\zeta, \zeta')$ component, $K(\zeta|\zeta')$ means the conditional probability that the tetrahedron $T_{j+1}$ takes the state $\zeta$, given that $T_{j}$ is in the state $\zeta'$.
This probability does not depend on the index $j$.
$P_N(\zeta_N|\zeta_0)$ can be expressed as 
\begin{align}
P_N(\zeta_N|\zeta_0) &= \sum_{\zeta_1\cdots\zeta_{N-1}}K(\zeta_N|\zeta_{N-1})K(\zeta_{N-1}|\zeta_{N-2})\cdots K(\zeta_1|\zeta_0)\nonumber\\
&= K^N_{\eta_{N}\eta_0}.
\end{align}

With the eigenvalue of $K$, $\lambda_{\gamma}$, and corresponding right and left eigenvectors as ${\mathbf u}^{\gamma}$ and ${\mathbf v}^{\gamma}$,
one can write
\begin{eqnarray}
P_N(\zeta_N|\zeta_0) = \sum_{\alpha=1}^{16}\lambda_{\gamma}^Nu^{\gamma}_{\zeta_N}v^{\gamma}_{\zeta_0}.
\end{eqnarray}

\subsection{Energy and related quantities}
By taking the charge representation, the internal energy of the system is given by
\begin{eqnarray}
E = \langle\mathcal{H}\rangle = \Bigl(\frac{1}{2} - J\Bigr)\sum_{p}\langle Q_p^2\rangle - J\sum_{\langle p,q\rangle}\langle Q_pQ_q\rangle.
\end{eqnarray}
To evaluate internal energy, the thermal expectation value of square charge, $\langle Q_p^2\rangle$, and nearest-neighbor
charge correlation, $\langle Q_pQ_q\rangle$ are required.
These quantities can be easily obtained from the equation (\ref{chargecorrelation}).
We can obtain specific heat $C$ by taking the numerical derivative of $E/N_{\rm site}$ in terms of temperature.
We also estimate $M_{Q.Q}$, defined with eq. (\ref{eq:MQQ}), from $\langle Q_pQ_q\rangle$.
These quantities are plotted in Fig.~\ref{fig:obsJ0.25}, together with the results of Monte Carlo simulation.

\subsection{Magnetic susceptibility}
The magnetic susceptibility, $\chi_z$, can be obtained from the spin correlation introduced in equation (\ref{spincorrelation}).
By using the anisotropy axis ${\mathbf d}_{\alpha}$ associated with each sublattice, $\alpha$, one can write
\begin{eqnarray}
\chi_z = \frac{1}{T}\sum_j\langle\eta_0\eta_j\rangle d^z_{\alpha_0}d^z_{\alpha_j}.
\end{eqnarray}

\section{Dynamical processes}
\label{dynamical_processes}

For convenience, we summarize the important dynamical processes and the energy barriers, $\Delta_m$, associated with each process.
$\Delta_m$ is defined as $\Delta_m = E^{\rm f}_m - E^{\rm i}_m$, where $E^{\rm f(i)}_m$ is the total energy of the system after (before)
the process $m$ takes place.

\noindent
(1) Creation of a ``vacuum pair" from the [111] saturated state by flipping a spin in the kagome plane
\begin{eqnarray}
\Delta_1 = -4-20J.
\label{barrier1}
\end{eqnarray}

\noindent
(2) Creation of another vacuum pair next to the initial pair created in process (1)
\begin{eqnarray}
\Delta_2 = -4-16J.
\label{barrier2}
\end{eqnarray}

\noindent
(3) Creation of a dipole ($=$ a pair of charges) in the Coulomb phase
\begin{eqnarray}
\Delta_3=4-4J.
\label{barrier3}
\end{eqnarray}

\noindent
(4) Separation of two charges (starting from a dipole created in (3))
\begin{eqnarray}
\Delta_4=-4J.
\label{barrier4}
\end{eqnarray}

\noindent
(5) Creation of a double charge pair by flipping a spin on a triangular plane from the [111] saturated state
\begin{eqnarray}
\Delta_5=12+12J.
\label{barrier5}
\end{eqnarray}

\noindent
(6) spin flip just after (5): the pair of single charge and double charge becomes a pair of vacuum tetrahedron and single charge
\begin{eqnarray}
\Delta_6=-8-20J.
\label{barrier6}
\end{eqnarray}

\noindent
(7) spin flip just after (6): the other pair of single charge and double charge becomes a pair of vacuum tetrahedron and single charge
\begin{eqnarray}
\Delta_7=-8-16J.
\label{barrier7}
\end{eqnarray}

\noindent
(8) spin flip just after (6) and (7): vacuum-charge paper to single charge-double charge pair
\begin{eqnarray}
\Delta_8=8+16J.
\label{barrier8}
\end{eqnarray}

\noindent
(9) spin flip just after process (1): creation of a double-single charge pair by flipping the spin in the triangular layer, 
\begin{eqnarray}
\Delta_9=8+12J.
\label{barrier9}
\end{eqnarray}

\begin{figure*}
\centering\includegraphics[width=0.75\textwidth]{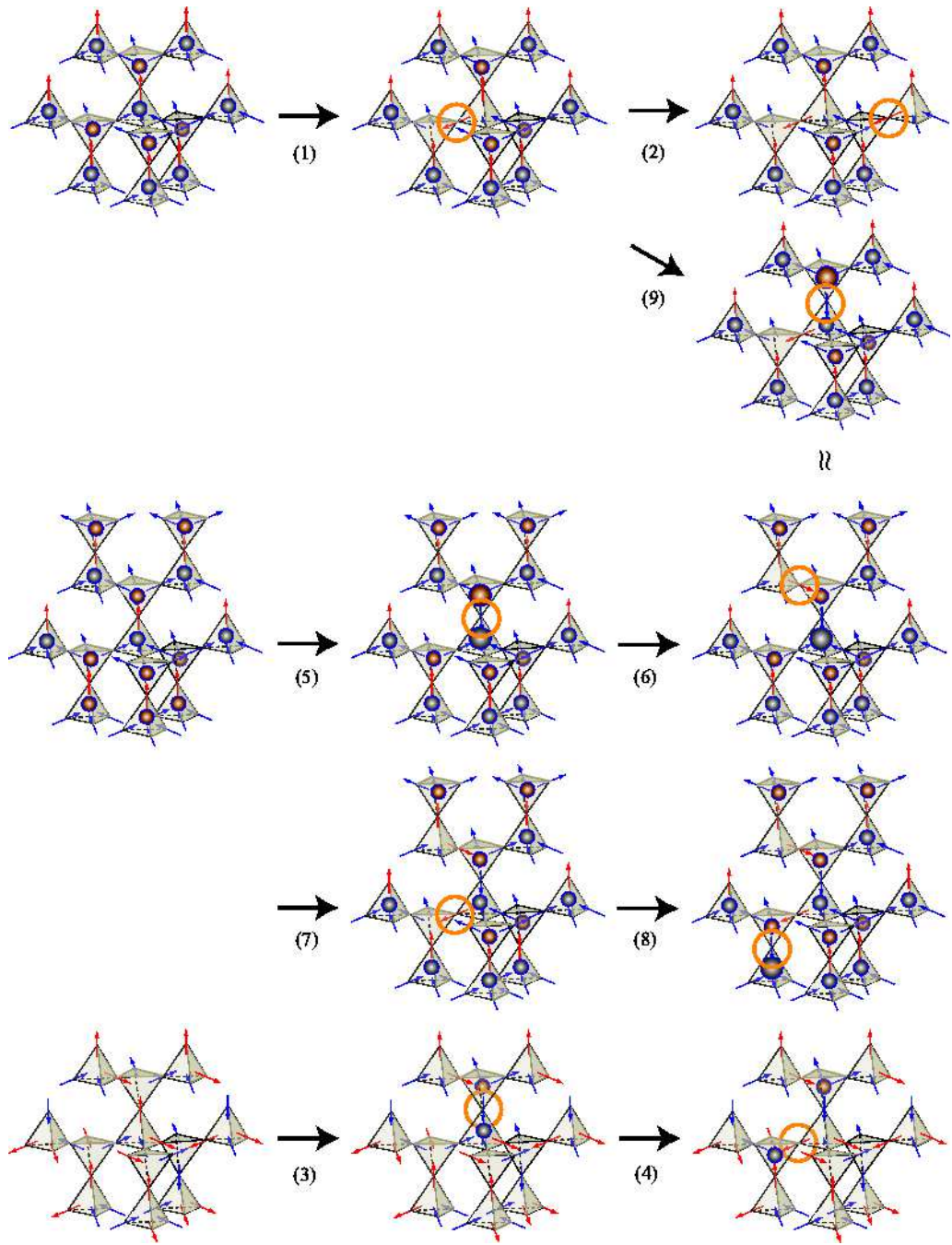}
\caption{\label{Fig_process1} 
(color online). Important dynamical processes. Each process is composed of a single spin flip. The flipped spin is highlighted with a yellow circle.
The energy barriers associated with each processes are listed as equations (\ref{barrier1}) to {(\ref{barrier9})}.
}
\end{figure*}

\section{Second plateau}
\label{appendix_plateau}

\begin{figure}[h]
\begin{center}
\includegraphics[width=0.45\textwidth]{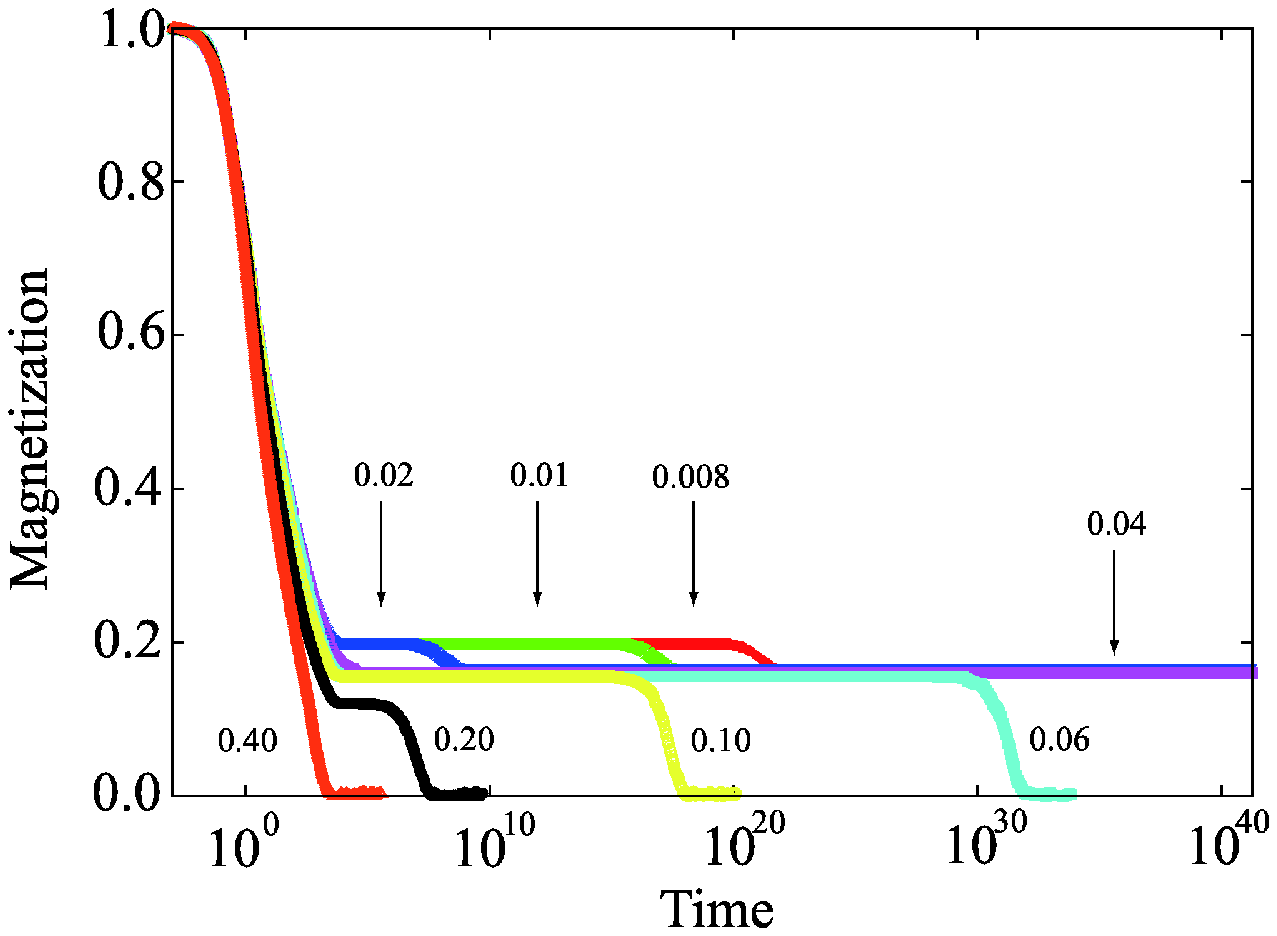}
\end{center}
\caption{\label{second_plateau} 
(color online). The time dependence of magnetization is plotted for $J=-0.10$ at $T=0.008, 0.01, 0.02, 0.04, 0.06, 0.10, 0.20$ and $0.40$.
At low temperatures, the magnetization shows the second plateau, at which $M\sim0.198$, 
which is slightly larger than the next plateau at $M\sim0.164$.
}
\end{figure}

For small negative $J$: $-0.2<J<0.0$, the magnetization exhibits a wide plateau until it is terminated due to 
the pair creation and dissociation processes of charges. Here, we note one additional feature appearing in the plateau region.
At low enough temperatures, the plateau splits into two regions, as shown in Fig.~\ref{second_plateau}.
The magnetization first drops to a value $M^*\sim 0.198$, then $M$ stays the same value for a certain time range.
After a while, the magnetization shows partial relaxation, and exhibits the second plateau.
After a long time determined from the pair creation process, the second plateau also collapses.
As temperature is raised, this first plateau shrinks, i.e., the collapse of first plateau occurs earlier at higher temperature,
and it finally merges with the initial drop of magnetization.

The origin of this first plateau can be attributed to the dissociation process of non-contractible charge pairs.
For negative $J$, charges with opposite sign attract with each other in the nearest-neighbor sites, which leads to the energy barrier $\Delta_4=4|J|$ for their dessociation.
Consequently, once two charges are trapped in a non-contractible position, it will take quite a long time to overcome the energy barrier to escape from that position. This mechanism is similar to that of
slow relaxation found in the dipolar spin ice model, where the energy barrier is attributed to the long-range dipolar interaction\cite{Castelnovo2010}.

\begin{figure}[h]
\centering\includegraphics[width=0.45\textwidth]{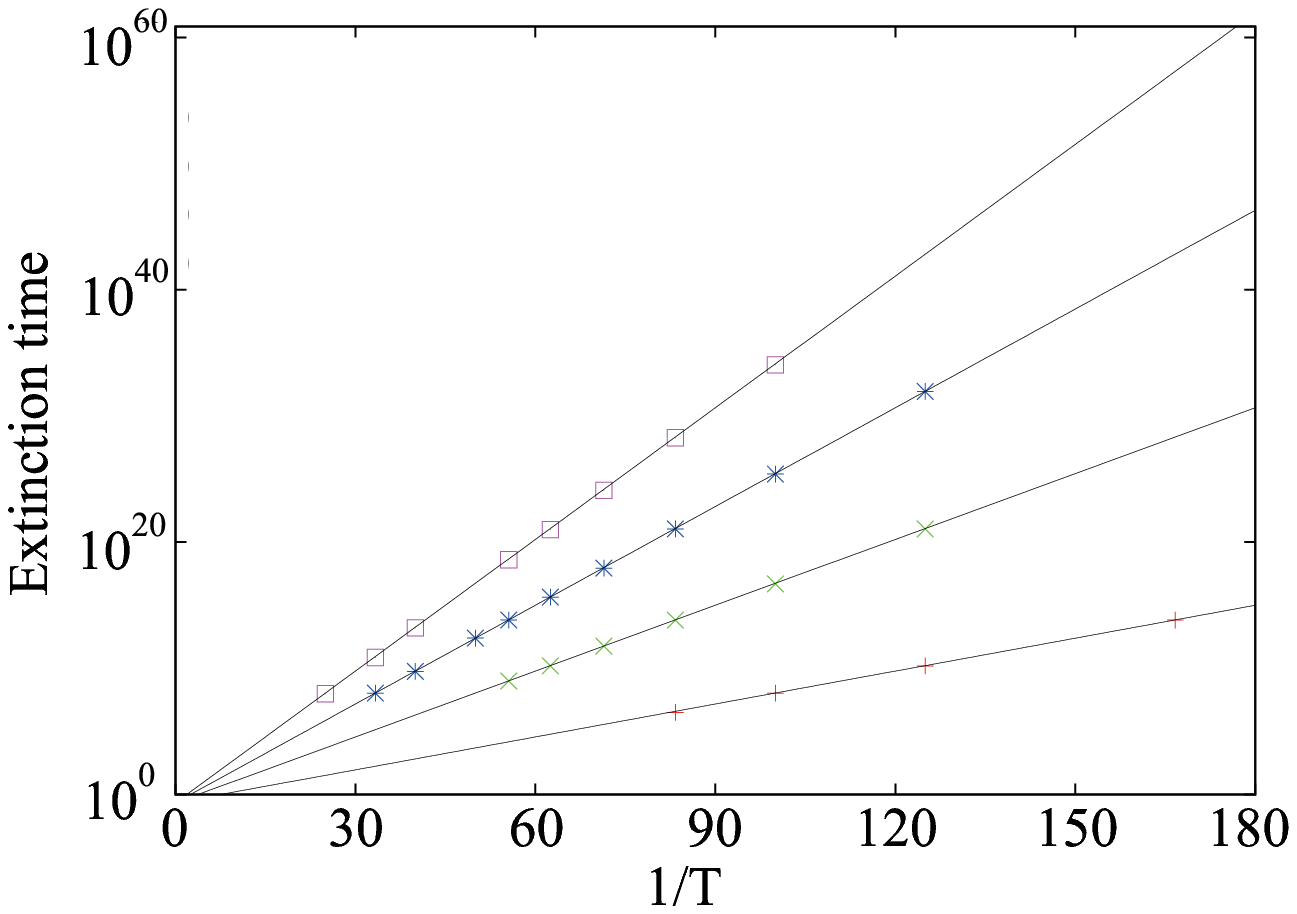}
\caption{\label{second_plateau_collapse} 
(color online). The time for the termination of first plateau is plotted for $J=-0.05, -0.10, -0.15$ and $-0.20$ from bottom to top. Each data are well fitted with Arrhenius-type temperature dependence: $\log t = -1.5 + 4J/T$, as shown above.}
\end{figure}

As plotted in Fig.~\ref{second_plateau_collapse}, the time necessary to terminate the first plateau is well fitted by an 
Arrehnius law: $\exp(\Delta_4/T)$, with the value of energy barrier associated with the dissociation process.

\section{Absence of Hall effect in the charge crystal}
\label{appendix_Hall}

\begin{figure}[h]
\begin{center}
\includegraphics[width=0.45\textwidth]{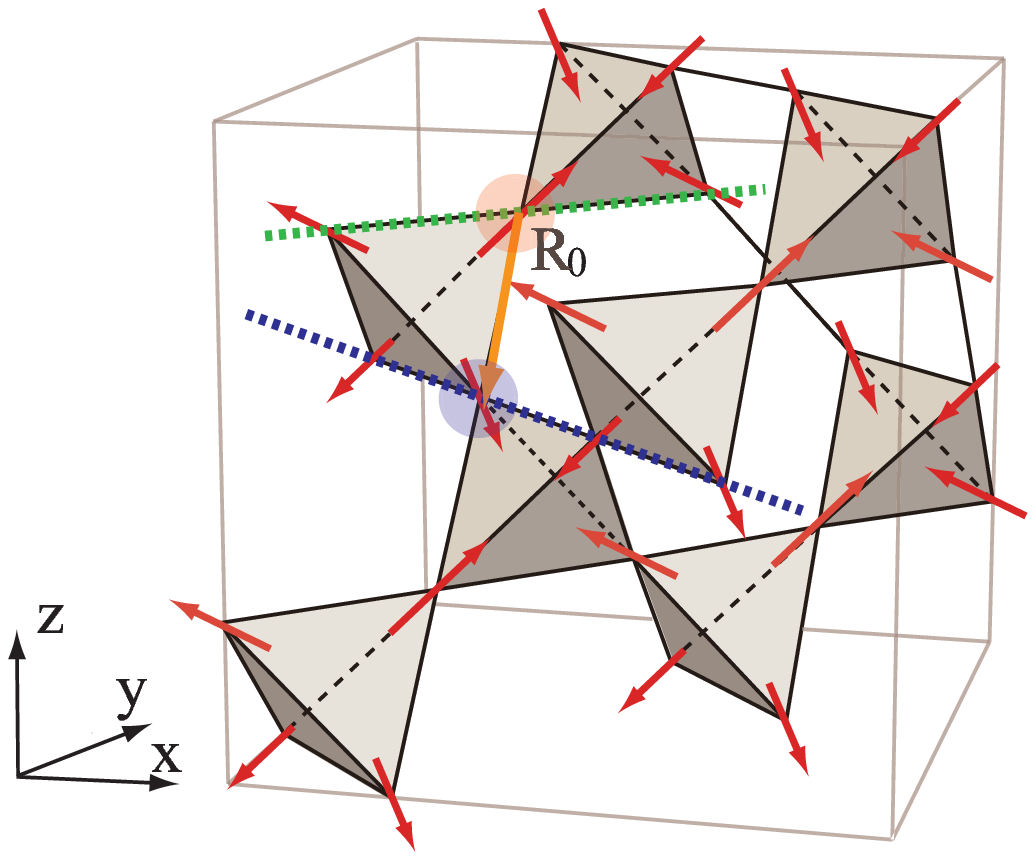}
\end{center}
\caption{\label{noHall_monopole_crystal} 
(color online). Schematic picture of the all-in crystal state. The green dashed line shows one of the $\beta$-chains, which will be transformed to the blue line after the operation $RT$ (see the main text).}
\end{figure}

We can show that the Hall conductivity vanishes for the FCSL, on the basis of spatial symmetry of this state. To prove this, it is instructive first to consider the case of all-in/all-out crystal [Fig.~\ref{noHall_monopole_crystal}]. Here, we define all-in (all-out) crystal as the state where all the upward tetrahedra take the all-in (all-out) configuration.

Firstly, the transverse conductivity $\sigma_{xy}$ satisfies the relation
\begin{eqnarray}
{\mathbf J}_y = \sigma_{xy}{\mathbf E}_x.
\end{eqnarray}
Here, ${\mathbf J}_y$ and ${\mathbf E}_x$ are electric current in the $y$ direction, and electric field in the $x$ direction.
The Hall conductivity $\sigma^{\rm H}_{xy}$ is defined as antisymmetric part of transverse conductivity $\sigma_{xy}$, as
\begin{eqnarray}
\sigma^{\rm H}_{xy} = \frac{\sigma_{xy} - \sigma_{yx}}{2}.
\end{eqnarray}
Accordingly, $\sigma^{\rm H}_{xy}$ satisfies 
\begin{eqnarray}
\sigma^{\rm H}_{xy} = -\sigma^{\rm H}_{yx},
\end{eqnarray}

On the basis of this property, we prove that Hall conductivity vanishes in the all-in/all-out ordered state.
To begin with, we note that the all-out state is obtained by reversing all the spin directions from the all-out state. In other words,
$T:$ Time-reversal operation maps the all-in state to the all-out state, and vice versa.
Next, we define the spatial operation $R$, by combining rotational and translational operations.
Firstly, we consider the $90^{\circ}$ rotation (of spin and position) around an line parallel to $z$-axis, and through one of the sites with sublattice $B$ [Fig.~\ref{noHall_monopole_crystal}]. Then, we successively translate the system by ${\rm R}_0$. This combined operation, $R$, maps a group of $\beta$-chains to those perpendicular to the former. 
At the same time, this operation interchanges the upward and downward tetrahedra, and accordingly, map the all-in state to the all-out state, and vice versa.
From the viewpoint of ${\mathbf J}_y$ and ${\mathbf E}_x$, the operation, $R$, acts as
\begin{eqnarray}
{\mathbf J}_y\to-{\mathbf J}_x, \hspace{1cm}{\mathbf E}_x\to{\mathbf E}_y.
\end{eqnarray}

\noindent
So, the combination of $T$ and $R$, maps the all-in and all-out states to themselves. Consequently, $TR$ (or $RT$) does not change the value of $\sigma_{xy}$. Meanwhile, $TR$ (or $RT$) map
\begin{eqnarray}
{\mathbf J}_y\to{\mathbf J}_x, \hspace{1cm}{\mathbf E}_x\to{\mathbf E}_y.
\end{eqnarray}
Accordingly, we have
\begin{eqnarray}
{\mathbf J}_y = \sigma_{xy}{\mathbf E}_x \to {\mathbf J}_x = \sigma_{xy}{\mathbf E}_y.
\end{eqnarray}
So, $\sigma_{yx} = \sigma_{xy}$, and consequently, $\sigma^{\rm H}_{xy}=0$.\\

The same proof holds, if the exchange of upward and downward tetrahedra can be considered equivalent to time-reversal operation. 
In the FCSL, the upward and downward tetrahedra are occupied with 3-in 1-out and 1-in 3-out configurations, or vice versa.
Accordingly, while the FCSL is spin-disordered state, on a macroscopic scale, the exchange of upward and downward tetrahedra lead to 
the time-reversal conjugate state of initial state, resulting in the absence of Hall signal in FCSL.

\section{Collapse of jellyfish}
\label{collapse_of_jellyfish}
The jellyfish structure is quite stable.
Accordingly, the toroidal moment associated with its ring part has quite a long lifetime.
Here, we discuss two dominant processes which collapse the ring structure and destroy the toroidal moment, accordingly.
Both processes lead to a lifetime of the order of at least $\sim\exp(J_1/T)=\exp(1/T)$.
Accordingly, this structure has a quite long lifetime at sufficiently low temperatures.

\subsection{energy barriers}
\subsubsection{double charge-vacuum creation}
\label{dn_creation}
Since the jellyfish is composed of the single charges with the same sign, it is subject to the creation of double charge-vacuum pair
from the adjacent two single charges. In particular, if it happens in the ring part, this event immediately disturbs the chiral magnetization flow.
The energy cost for this pair creation, $\Delta_{\rm p-c}$, is estimated as $\Delta_{\rm p-c} = 4(1-J)$, if the jellyfish has no branches.
In case the branches take the optimal configuration for the collapse, it is reduced to $\Delta_{\rm p-c}^{\rm min} = 4(1-3J)$, but still this lower-bound value takes $\sim1$ near $J=1/4$, so this decay process hardly occurs at $T\ll1$. 

\begin{figure}[t]
\begin{center}
\includegraphics[width=0.45\textwidth]{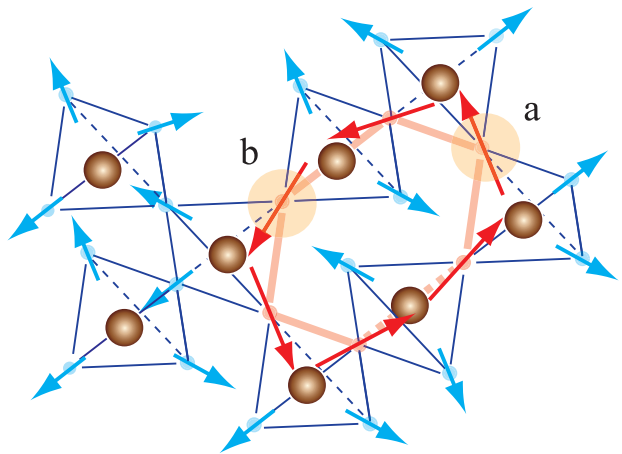}
\end{center}
\caption{\label{pc_process1} 
(color online) The schematic figure of jellyfish with two branches, each of which has one charge, respectively. The magnetization flow is shown with red arrows. The reversal of spin $a$ costs the energy $\Delta_{\rm p-c} = 4(1-J)$. Meanwhile the energy cost of reversing spin $b$ takes the lowest possible value, $\Delta_{\rm p-c}^{\rm min}=4(1-3J)$.}
\end{figure}

\subsubsection{pair annihilation of opposite charges}
\label{pair_annihilation}
Pair annihilations of opposite charges is another possible process leading to the decay of jellyfish.
Due to the topological constraint, the total charges in the system must be equal to zero.
Accordingly, if there is a jellyfish composed of a certain numbers of positive charges, there must be the same numbers of negative charges elsewhere in the system, which is likely to form a jellyfish or methane cluster to lower their energy, near $J=1/4$.
Under the circumstances, to destroy the positive jellyfish, one can consider the following three-step process:
(i) one monopole is detached from the negative jellyfish/methane, (ii) it migrates to the positive jellyfish, and (iii) makes a pair-annihilation with one of the positive charges composing the target positive jellyfish.

The energy barrier accompanying (i) is at least $4J$. (iii) also leads to the energy barrier of $4J$, since it requires the opposite charges to be placed in nearest-neighbor sites. So the total energy cost for this pair-annihilation, $\Delta_{\rm p-a}$, is estimated as $\Delta_{\rm p-a} = 8J$.
However, if there are finite numbers of stray charges, possibly due to thermal excitation, the step (i) is not necessary. In this case,
$\Delta_{\rm p-a}$ is reduced to $\Delta_{\rm p-a}^{\rm min}=4J$, and it gives a comparable value to the lower bound of double charge-vacuum creation energy barrier: $\Delta_{\rm p-a}^{\rm min}\sim\Delta_{\rm p-c}^{\rm min}\sim1$, near $J=1/4$.

\subsection{kinematic constraint}
In addition to the energy barriers discussed above, there are substantial contributions from the kinematic constraint.
Firstly, the double charge-vacuum creation costs the energy $\Delta_{\rm p-c}\geq\Delta_{\rm p-c}^{\rm min}=4(1-3J)$. However, the lower-bound value, $\Delta_{\rm p-c}^{\rm min}$ is available, only if a special configuration is realized on a ring [see Fig.~\ref{pc_process1}], which is statistically rare.

Secondly, the lifetime estimated from the pair-annihilation process also requires substantial kinematic correction. This process involves the migration of charge, which we call step (ii) above. If the charge density is low, the charge has to migrate for quite a long time until it successfully go overs the energy barrier due to the step (iii).

These kinematic features contrast with the dynamics at $J<0$, where the dynamical bottleneck is given by a single energy barrier in most cases, and the dynamics after overcoming the barrier is avalanche-like.
The kinematic constraints give a substantially large correction to the lifetime $\sim\exp(1/T)$ estimated purely from energetic consideration,
and contributes to the stability of jellyfish structures.

\begin{figure}[t]
\begin{center}
\vspace{0.2cm}
\includegraphics[width=0.45\textwidth]{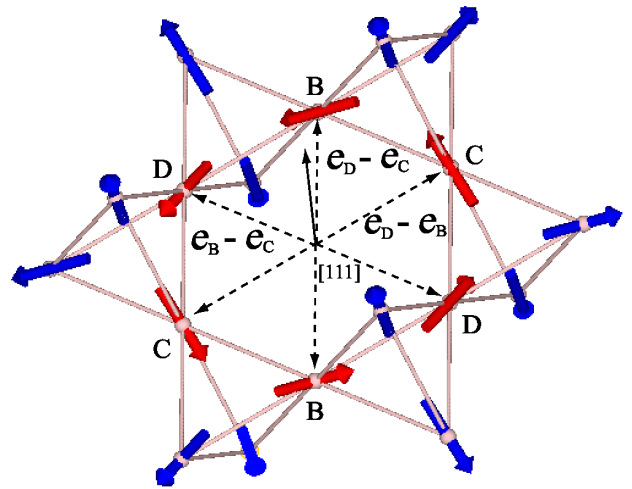}
\end{center}
\caption{\label{monopole_ring_structure} 
(color online) The schematic picture of charge ring, perpendicular to $[111]$. For the sites on the hexagonal ring, the sublattice indices and the relative positions from the center of the hexagon are shown.
}
\end{figure}

\section{Effects of charge rings on structure factor}
\label{ring_to_SofQ}
To consider how the jellyfish affects the magnetic structure factor, we evaluate the 
contribution from the ring part of jellyfish to the spin correlation function. Namely, we calculate
\begin{eqnarray}
\mathcal{S}_{\rm ring}({\mathbf q}) = \langle{\mathbf S}^{\rm ring}_{\mathbf q}\cdot{\mathbf S}^{\rm ring}_{-{\mathbf q}}\rangle,
\end{eqnarray}
where
\begin{eqnarray}
{\mathbf S}^{\rm ring}_{\mathbf q} = \sum_{i\in{\rm ring}}{\mathbf S}_ie^{-i{\mathbf q}\cdot{\mathbf r}_i}.
\label{def_Sq_ring}
\end{eqnarray}

The ring part can be classified into four types, according to its orientation: the hexagonal ring is perpendicular to [111], [1-1-1], [-11-1] and [-1-11].
We assume these four types of rings are equally populated, without any correlations on their positions. For each ring, we take account of the contribution from the 18 site-cluster, as shown, in Fig.~\ref{monopole_ring_structure}. For example, from the ring perpendicular to [111], we obtain
\begin{widetext}
\begin{align}
{\mathbf S}^{[111]}_{\mathbf q} &\propto\eta_{\rm ch}[{\mathbf d}_{\rm B}\sin({\mathbf q}\cdot({\mathbf e}_{\rm C} - {\mathbf e}_{\rm D})) + {\mathbf d}_{\rm C}\sin({\mathbf q}\cdot({\mathbf e}_{\rm D} - {\mathbf e}_{\rm B})) + {\mathbf d}_{\rm D}\sin({\mathbf q}\cdot({\mathbf e}_{\rm B} - {\mathbf e}_{\rm C}))]\nonumber\\
&+ \eta_{\rm pn}[{\mathbf d}_{\rm B}\sin({\mathbf q}\cdot({\mathbf e}_{\rm C} + {\mathbf e}_{\rm D} - 2{\mathbf e}_{\rm B})) + {\mathbf d}_{\rm C}\sin({\mathbf q}\cdot({\mathbf e}_{\rm D} + {\mathbf e}_{\rm B} - 2{\mathbf e}_{\rm C})) + {\mathbf d}_{\rm D}\sin({\mathbf q}\cdot({\mathbf e}_{\rm B} + {\mathbf e}_{\rm C} - 2{\mathbf e}_{\rm D}))\nonumber\\
&\hspace{0.2cm}+{\mathbf d}_{\rm A}\{\sin({\mathbf q}\cdot({\mathbf e}_{\rm B} + {\mathbf e}_{\rm C} - {\mathbf e}_{\rm D} - {\mathbf e}_{\rm A})) + \sin({\mathbf q}\cdot({\mathbf e}_{\rm C} + {\mathbf e}_{\rm D} - {\mathbf e}_{\rm B} - {\mathbf e}_{\rm A})) + \sin({\mathbf q}\cdot({\mathbf e}_{\rm D} + {\mathbf e}_{\rm B} - {\mathbf e}_{\rm C} - {\mathbf e}_{\rm A}))\}].
\label{Sqfrom111}
\end{align}
\end{widetext}
Here, ${\mathbf e}_{\{i={\rm A, B, C, D}\}}$ means the internal coordinate of sublattice A, B, C, D within a unit cell.
The first term in (\ref{Sqfrom111}) comes from the hexagonal ring, and $\eta_{\rm ch}=\pm1$ determines the chirality of the spins. The other terms come from the outer part, and $\eta_{\rm pn}=\pm1$ means the sign of constituent charges.

${\mathbf S}^{\rm ring}_{\mathbf q}$ for other orientations of rings on [1-1-1], [-11-1] and [-1-11] can be obtained by
replacing the indices in equation (\ref{Sqfrom111}) as (ABCD)$\to$(BADC), (CABD), (DACB), respectively. The summation:
\begin{eqnarray}
\mathcal{S}_{\rm ring}({\mathbf q}) = \mathcal{S}^{[111]}_{\mathbf q} + \mathcal{S}^{[1-1-1]}_{\mathbf q} + \mathcal{S}^{[-11-1]}_{\mathbf q} + \mathcal{S}^{[-1-11]}_{\mathbf q}.
\end{eqnarray}
is proportional to the total contribution from the rings on structure factor.


\end{document}